\documentclass[doublecol]{epl2}

\usepackage{amsmath,color}
\usepackage{mymacros}

\title{Turbulent radial thermal counterflow in the framework of the HVBK model}

\author{Y.\,A.~Sergeev \and C.\,F.~Barenghi}
\shortauthor{Y.\,A.~Sergeev and C.\,F.~Barenghi}

\institute{                    
  Joint Quantum Centre Durham-Newcastle, and School of Mathematics, Statistics and Physics, Newcastle University, Newcastle upon Tyne, NE1 7RU, United Kingdom
}
\pacs{47.37.+q}{Hydrodynamic aspects of superfluidity}
\pacs{67.25.dk}{Vortices and turbulence}
\pacs{67.25.dm}{Two-fluid model: phenomenology}

\abstract{We apply the coarse-grained Hall-Vinen-Bekarevich-Khalatnikov (HVBK) equations to model the statistically steady-state, turbulent, cylindrically symmetric radial counterflow generated by a moderately large heat flux from the surface of a cylinder immersed in superfluid $^4$He. We show that a time-independent solution exists only if a spatial non-uniformity of temperature and the dependence on temperature of the thermodynamic properties are accounted for. We demonstrate the formation of a thermal boundary layer whose thickness grows with temperature of the cylinder's surface, and analyze the properties of the flow in the radial
direction, including the local average vortex line density.}

\begin{document}

\maketitle

\section{Introduction}
\label{sec:intro}

We develop a phenomenological model for the statistically steady-state quantum (superfluid) turbulence generated by a heat flux from a surface of the infinitely long cylinder immersed in superfluid $^4$He. This flow is hereafter referred to as the turbulent radial thermal counterflow. At temperatures between $\sim0.7\,{\rm K}$ and the superfluid transition temperature $\Tl\approx2.17\,{\rm K}$ liquid helium can be modelled as an intimate mixture of inviscid superfluid and viscous normal fluid. Microscopically, quantum turbulence manifests itself in the superfluid 
component of liquid helium as a dynamic tangle of interacting quantized vortex lines which move around each others and reconnect when they collide. Macroscopically, the intensity of quantum turbulence is characterized by the vortex line density, $L$ defined as the length of quantized vortex lines per unit volume larger than the average vortex separation $\ell\approx L^{-1/2}$. In many situations, the vortex line density can be considered as a statistically steady state quantity. In the termal counterflow (which has no analogue in classical fluid dynamics) the flow of normal fluid can be either laminar, or turbulent.

In the dense turbulent counterflow considered below, for $L$ ranging from $10^7$ to $10^{12}\,{\rm cm}^{-2}$ (see below), the typical intervortex spacing $\ell$ ranges from $10^{-6}$ to $10^{-3}$~cm. Simple estimates show that the Reynolds number defined by $\ell$ is small so that the normal flow is laminar (T1 regime of quantum turbulence~\cite{Tough}) for all physically meaningful values of parameters.

Assuming that the temperature of the cylinder's surface is $\To$ and that the heat flux generated by the surface is $q$, we analyze the radial distributions of temperature and of the normal, superfluid, and counterflow velocities ($\vvn$, $\vvs$, and $\vvns=\vvn-\vvs$, respectively) taking into account the dissipation caused by the mutual friction, as well as the changes of normal and superfluid densities ($\rhon$ and $\rhos$, respectively) and other thermodynamic properties of helium with temperature. The simplest self-consistent model that links the temperature distributions with the flow properties is based on the Hall-Vinen-Bekarevich-Khalatnikov (HVBK) equations which are discussed below.

Heat transfer between helium II and the wire, heated by an applied electric current, is an essential feature of experiments~\cite{Shiotsu,Ruzhu,Duri} aimed at developing hot-wire anemometry (a standard technique of fluid mechanics) to turbulent $^4$He. The diameters of the wires used in the
experiments~\cite{Shiotsu,Ruzhu} ranged from 50 to 80~$\mu$m, while much thinner probes of diameter 1.3~$\mu$m have been used in other experiments~\cite{Duri}. In these experiments the wire was overheated to temperatures up to 25~K by typical heat fluxes of the order of $100\,{\rm W/cm^2}$. Dur\`i et al.~\cite{Duri} identified three regions in the liquid helium surrounding the wire, each with its own distinct mechanism of heat transfer: (a)~A cylindrical shell of thickness about 0.2~$\mu$m adjacent to the wire where the liquid helium is in the supercritical normal phase; 
in this region the temperature decreases from $\To$ to $\Tl$ with the distance from the wire's surface, and heat transfer is due to the molecular conductivity. (b)~A thermal boundary layer of thickness about a tenth of the
wire's diameter adjacent to the first layer, where the molecular conduction is complemented by a radial thermal counterflow; in this region the temperature drops with distance from $\Tl$ to a somewhat lower value. (c)~An outer region where the heat transfer is dominated by the turbulent radial counterflow. Note that the combined thickness of the first two zones do not exceed 15~--~20\% of the wire's diameter. Dur\`i et al.~\cite{Duri} analyzed the temperature distribution in region (a) by numerical integration of Fourier's law, and in regions (b) and (c) by numerical analysis of a simple model based on the Gorter-Mellink's semi-empirical correlation (see e.g. Ref.~\cite{VanSciver-book}) for the overall mean heat flux in He~II. Although they did not analyze the distribution of the local vortex line density around the hot-wire in detail, Dur\`i et al. noted that a large heat flux and a temperature close to $\Tl$ near the wire surface lead to very high counterflow velocities, of the order of the second sound speed in some cases. Such a counterflow should produce a very dense vortex tangle in the close vicinity of the wire; their estimates~\cite{Duri} suggest that the mean intervortex spacing might be as small as 0.01~$\mu$m yielding a local vortex line density  up to $L \approx 10^{12}\,{\rm cm^{-2}}$ in the vicinity of the wire.

The aim of this work is not to consider the regions (a) and (b) of the heat transfer, but analyze in more detail region (c), where the turbulent radial counterflow is generated by the heated cylinder maintained at a temperature $\To$ that is somewhat lower than $\Tl$. In polar cylindrical coordinates, the normal and superfluid velocities, whose only non-zero components are in the radial direction, the normal and superfluid densities ($\rhon$ and $\rhos$, respectively), and the thermodynamic properties are functions of the radial coordinate $r$ alone. At the surface of cylinder, $r=a$, where $a$ is the cylinder's radius, the normal velocity, $\vn(r)$, and the counterflow velocity, $\vns(r)=\vn(r)-\vs(r)$, are linked to the heat flux from the cylinder's surface by the relations
\begin{equation}
\vo=\vn(a)=\frac{q}{\rho s_0\To}\,, \quad \vns(a)=\frac{q}{\rhos(\To)s_0\To}\,,
\label{eq:q}
\end{equation}
where $\rho=0.145\,{\rm g/cm^3}$ is the total helium's density which we assume is temperature-independent, $s(T)$ is the entropy per unit mass, and $s_0=s(\To)$.

Our model is based on the HVBK equations~\cite{Khalatnikov-book} for the non-isothermal flow of turbulent $^4$He (see also Refs.~\cite{Nemirovskii1995,Donnelly1999} for the comprehensive review of the HVBK equations). We expect our model to describe, at least qualitatively, some (albeit not all) of the features of radial turbulent counterflow similar to that discussed in Refs.~\cite{Shiotsu,Ruzhu,Duri}. 

Closely related to our model developed below are the study~\cite{Kafkalidis} of radial turbulent counterflow in a diverging channel\footnote{In experiment~\cite{Kafkalidis} spatial variations of temperature did not exceed 30~mK, so that, unlike the present study, the mathematical model developed in Ref.~\cite{Kafkalidis} ignored variations with temperature of the normal and superfluid densities and thermodinamic properties.}, and a recent numerical simulation\cite{Varga} of spherically symmetric counterflow. In the latter work, the temperature of helium was assumed to be uniform throughout the flow domain, and the simulations were implemented using the vortex filament method with the full Biot-Savart interactions and algorithmic vortex reconnections.

\section{HVBK equations, mutual friction, and evolution of the vortex line density}
\label{sec:HVBK}

For non-isothermal flows of turbulent superfluid helium, the HVBK equations are
\begin{align}
&\pd{\rho}{t}+\bnabla\cdot\jj=0\,, \label{eq:mass-cons} \\
&\pd{s}{t}+\bnabla\cdot(s\vvn+\bSigma)=\frac{R}{T}\,, \label{eq:entropy} \\
&\pd{\jj}{t}+\bnabla\cdot\bPi={\bf 0}\,,\label{eq:flux} \\
&\rhos\left[\pd{\vvs}{t}+(\vvs\cdot\bnabla)\vvs\right] \nonumber \\
&=-\frac{\rhos}{\rho}\bnabla p_s+\rhos s\bnabla T+{\bf f}\,, \label{eq:s_momentum}
\end{align}
where $\jj=\rhos\vvs+\rhon\vvn$ is the total momentum density, $\bSigma$ is the entropy flux associated with the presence of vortex tangle, $R$ is the rate of heat production by the mutual friction and reactive forces, $\bPi$ is the momentum flux density tensor whose components are
\begin{align}
\Pi_{ij}&=\rhos\vs_i\vs_j+\rhon\vn_i\vn_j+p\delta_{ij} \nonumber \\
&-\eta\left(\pd{\vn_i}{x_j}+\pd{\vn_j}{x_i}-\frac{2}{3}\delta_{ij}\bnabla\cdot\vvn\right)
\label{eq:tensor}
\end{align}
(here $x_k$ are spatial coordinates, $p$ is the pressure, $\delta_{ij}$ is the Kronecker delta, and $\eta$ is the temperature-dependent viscosity of the normal fluid), the gradient of effective pressure, $p_s$ acting on the superfluid component is $\bnabla p_s=\bnabla p-\tfrac{1}{2}\rhon\bnabla\vns^2$, and ${\bf f}$ is the force acting on the superfluid component.

We model the thermal counterflow generated by a rather large heat flux so that the radial normal velocity at the cylinder's surface, $\vo$,
can be up to $\sim100\,{\rm cm/s}$ (note that the typical heat flux in experiments~\cite{Duri} could be up to a few hundred ${\rm W/cm^2}$). We assume that the surface temperature can be up to 2.15~K. At higher temperatures in the interval between 2.15~K and $\Tl$ the temparure derivatives of thermodynamic properties become too large; furthermore, as $T\to\Tl$, $\rhos\to0$, $\vs\to\infty$ and $\vns\to\infty$. This results in unphysically large vortex line densities and infinite gradients of the flow properties, so that the coarse-grained HVBK model breaks down\footnote{Our calculations predict rather large gradients of temperature and vortex line density in the case where the surface temperature is sufficiently close to 2.15~K. The validity of the HVBK model in the case where the vortex tangle is strongly non-uniform is yet unclear. Here we adopt a modeling approach, and assume that even in the case of strong non-uniformity, the HVBK-based calculations are at least qualitatively correct.}.

For a detailed discussion of closures for $\bSigma$, $R$, and ${\bf f}$ we refer the reader to Refs.~\cite{Nemirovskii1995,Donnelly1999,Nemirovskii1983}. Here it suffices to say that, for large counterflow and superfluid velocities and large vortex line densities typical of examples considered in this paper, the direct estimates of $\bSigma$ and $R$ show that their contribution can be neglected, and that ${\bf f}$ can be approximated by the mutual friction force in the Gorter-Mellink form
\begin{equation}
{\bf f}\approx {\bf F}_{ns}=\rhos\alpha\kappa L\vvns\,,
\label{eq:f}
\end{equation}
where $\kappa=0.997\times10^{-3}\,{\rm cm^2/s}$ is the quantum of circulation, 
and $\alpha=\alpha(T)$ is the temperature-dependent mutual friction coefficient tabulated in Ref.~\cite{Donnelly-Barenghi}. 
The vortex line density is regarded here as a locally averaged quantity, $L=L({\bf r},\,t)$, whose evolution obeys Vinen's equation~\cite{Vinen} modified~\cite{Nemirovskii1995,Nemirovskii1983} 
to account for the tangle's drift in the non-uniform quantum turbulence:
\begin{equation}
\pd{L}{t}+\bnabla\cdot(L{\bf v}_L)=\alpha_V\vert\vvns\vert^2L^{3/2}-\beta_VL^2\,,
\label{eq:Vinen}
\end{equation}
where ${\bf v}_L=b(T)\vvns$, with $b(T)$ being a small, non-dimensional, temperature-dependent coefficient (for temperatures between 1.3 and 2~K calculations of Ref.~\cite{Kondaurova} yield the values of $b$ between 0.04 and 0.1), $\alpha_V$ and $\beta_V$ are temperature-dependent coefficients\footnote{Based on the further generalization of Vinen's equation which also included the term responsible for the diffusion of the local vortex line density, a theoretical analysis of radial counterflow in the case of small temperature gradient has been performed in Ref.~\cite{Saluto2014}.}.

Furthermore, we consider a steady-state regime and neglect the relatively slow drift of the vortex tangle (later we shall justify this approximation using estimates obtained from our numerical solution). Then Eq.~(\ref{eq:Vinen}) reduces to the following well-known relation which closes the system of equations~(\ref{eq:mass-cons})-(\ref{eq:entropy}) and (\ref{eq:s_momentum}):
\begin{equation}
L=\gamma^2\vert\vvns\vert^2\,,
\label{eq:gamma}
\end{equation}
where $\gamma$ is a temperature-dependent parameter. The values of $\gamma$ at several temperatures are found both from numerical simulations~\cite{Adachi,Kondaurova} and from the counterflow experiments~(see e.g. Ref.~\cite{Babuin} and references therein; a comprehensive review of numerical and experimental evaluations of $\gamma$ can be found in Ref.~\cite{Kondaurova}). In applying our model to a steady-state radial counterflow, we must assume that relation~(\ref{eq:gamma}) is still applicable in a situation where radial gradients of the vortex line density are large (e.g. at small distances from the heated cylinder), yielding at least qualitatively correct results. [Note that in Eq.~(\ref{eq:gamma}) we also neglected the so-called intercept velocity~\cite{Tough} which, being of the order of 0.1~cm/s, is much smaller than typical values of $\vns$ in our calculations.]

We also neglect the viscous stress term in the momentum flux density tensor~(\ref{eq:tensor}); estimates based on our numerical solutions show that the contribution of the viscous term to the momentum transfer in the normal fluid is much smaller than that of the mutual friction term even in the region where the gradients of flow properties are large.

For the steady-state turbulent counterflow, the only non-zero components of the velocities $\vvn$, $\vvs$, and $\vvns$ are their radial components $\vn$, $\vs$, and $\vns$ (hereafter we omit the subscript which indicates the radial direction). These quantities, as well as the temperature, $T$, 
and the local average line density $L$, are functions of the radial coordinate $r$ alone, and $\rhos=\rhos(T)$, $\rhon=\rhon(T)$, $s=s(T)$, $\alpha=\alpha(T)$, and $\gamma=\gamma(T)$. From Eqs.~(\ref{eq:mass-cons}) and (\ref{eq:entropy}), neglecting the contributions of $\bSigma$ and $R$, for the steady-state counterflow ($\partial/\partial t\equiv0$) we have
\begin{equation}
\vn=\frac{as_0\vo}{rs}\,, \quad \vs=-\frac{\rhon}{\rhos}\vn, \quad \vns=\frac{\rho}{\rhos}\vn\,.
\label{eq:v-rad}
\end{equation}

Before proceeding with finding the radial profiles of temperature, velocities, and vortex line density, we  make some comments on the steady-state turbulent counterflow at uniform temperature.

\section{Steady-state, radial, turbulent counterflow at uniform temperature}
\label{subsec:T-const}

In the case where $T=\To={\rm constant}$, the densities, entropy, and mutual friction parameters have constant values throughout the flow domain [$s=\so=s(\To)$, $\rhon=\rhon(\To)$, $\rhos=\rhos(\To)$, etc.], so that the first of relations~(\ref{eq:v-rad}) simplifies to $\vn=a\vo/r$. For $\partial/\partial t\equiv0$, eliminating the pressure $p$ between Eqs.~(\ref{eq:flux}) [with $\bPi$ defined by relation~(\ref{eq:tensor})] and (\ref{eq:s_momentum}), and incorporating Eq.~(\ref{eq:gamma}) we obtain
\begin{equation}
\oder{\,\,\,}{r}(\vn\vns)=-\frac{\rho}{\rhon}\kappa\alpha\gamma^2\vns^3\,,
\label{eq:steady-state-T-const}
\end{equation}
where now $\alpha=\alpha(\To)$ and $\gamma=\gamma(\To)$. Substituting relations~(\ref{eq:v-rad}) yields
\begin{equation}
a\vo=\frac{2}{\kappa\alpha\gamma^2}\frac{\rhon\rhos^2}{\rho^3}\,.
\label{eq:vo}
\end{equation}
From Eq.~(\ref{eq:vo}) it follows that for any given uniform temperature, a steady-state, radial turbulent counterflow exists only for a single, fixed value of $\vo$ determined by Eq.~(\ref{eq:vo}) (that is, for a single value of the heat flux)\footnote{For example. for $T=1.9\,{\rm K}$, using the tables~\cite{Donnelly-Barenghi} and the value of $\gamma$ calculated in Ref.~\cite{Adachi}, we find $a\vo\approx0.07\,{\rm cm^2/s}$.}. Note that similar arguments for a spherically symmetric counterflow suggest that a steady-state solution does not exist for any value of $\vo$. This is consistent with numerical simulations\cite{Varga} which showed the absence of saturation of 
the vortex tangle to a time-independent state.

We conclude that the radial non-uniformity of the temperature distribution is an essential feature of any steady-state, turbulent, radial counterflow, see the next Section.

\section{Non-isothermal, steady-state, turbulent counterflow. Results}
\label{sec:T-distributed}

We return to Eqs. (\ref{eq:mass-cons})-(\ref{eq:f}) and (\ref{eq:gamma}), with $\partial/\partial t\equiv0$ and the radial velocities determined by relations~(\ref{eq:v-rad}), assuming now that the temperaure is not constant but $T=T(r)$, and that the densities $\rhon$ and $\rhos$, the entropy $s$, mutual friction coefficient $\alpha$, and the coefficient $\gamma$ are known functions of temperature. We leave the temperature dimensional, but scale the radial coordinate, densities, velocities, and entropy, respectively, as follows:
\begin{equation}
x=\frac{r}{a}\,, \quad r_{n,s}=\frac{\rho_{n,s}}{\rho}\,, \quad V_{n,s,ns}=\frac{v_{n,s,ns}}{\vo}\,, \quad S=\frac{s}{\sll}\,,
\label{eq:non-dim}
\end{equation}
where $\sll=1.583\times10^7\,{\rm erg/(g\,K)}$ is the entropy at the $\lambda$-point. We also scale $\gamma(T)$ by its value calculated~\cite{Adachi} at $T=2.1\,{\rm K}$: $\gamma(T)=\gamma_{21}\Gamma(T)$, where $\gamma_{21}=\gamma(2.1\,{\rm K})=157\,{\rm s/cm^2}$. Eliminating the pressure $p$ between Eqs.~(\ref{eq:flux}), (\ref{eq:tensor}) and (\ref{eq:s_momentum}), incorporating Eqs.~(\ref{eq:f}) and (\ref{eq:gamma}), making use of the non-dimensionalized relations~(\ref{eq:v-rad}), and taking the temperature dependence of the normal and superfluid densities, thermodynamic properties and mutual friction parameters, we obtain, after rather lengthy algebra, the following equation for the radial distribution of temperature:
\begin{equation}
\left[1-\frac{\beta}{x^2}F(T)\right]\oder{T}{x}=\frac{1}{x^3}[\beta G(T)-\sigma H(T)]\,,
\label{eq:T(x)}
\end{equation}
where
\begin{align}
&F(T)=\frac{1}{2}\frac{\rn}{\rs S^3}\left(\frac{1}{\rs}\oder{\rs}{T}+\frac{2C_s}{ST}\right)\,,
\label{eq:F-C} \\
&G(T)=\frac{\rn}{\rs S^3}\,, \quad H(T)=\frac{\alpha\Gamma^2}{S^4\rs^3}\,, \label{eq:GH}
\end{align}
and the parameters $\beta$ and $\sigma$, which have the dimension of temperature, are
\begin{equation}
\beta=\so^2\vo^2/\sll^3\,, \quad \sigma=\kappa\so^3\vo^3\gamma_{21}^2a/\sll^4\,.
\label{eq:beta}
\end{equation}
To derive Eq.~(\ref{eq:T(x)}), we made use of the relation ${\rm d}s/{\rm d}T=c_s/T$ between the entropy, $s$ and the heat capacity, $c_s(T)$ (tabulated in Ref.~\cite{Donnelly-Barenghi}). In Eq.~(\ref{eq:F-C}), the heat capacity has been non-dimensionalized using the scaling $C_s=c_s/\sll$. 

For typical radii, $a$ in the range from $10^{-4}$ to $10^{-1}$~cm, and the velocity $\vo$ not larger than 150~cm/s, the value of $\beta$ is in the interval from 0 to $\sim1.5\times10^{-3}\,{\rm K}$, and $\sigma$ ranges from 0 to $\sim0.2\,{\rm K}$. Note that in Eq.~(\ref{eq:T(x)}), notwithstanding rather small values of $\beta$, the terms containing $\beta$ cannot be neglected because at temperatures sufficiently close to $\Tl$ (for instance at $T=2.15$) the values of $F(T)$ and $G(T)$ become very large. Functions $F(T)$, $G(T)$, and $H(T)$ are illustrated in Supplementary Material~\cite{Suppl}. Note that for all physically meaningful values of parameters $\beta G(T)<\sigma H(T)$ [in fact, in most cases $\beta G(T)\ll\sigma H(T)$], and also $1-\beta F(T)=O(1)$.

Equation~(\ref{eq:T(x)}) also enabled us to derive a sufficient condition for the bulk temperature to be regarded as practically uniform throughout the whole flow domain. In dimensional variables, such a condition has the form
\begin{equation}
a^{1/3}\vo\ll\frac{\rhos(\To)}{\rho}\left[\frac{\To s(\To)}{\kappa\alpha(\To)\gamma^2(\To)}\right]^{1/3}\,.
\label{eq:uniform}
\end{equation}
The details of derivation and analysis can be found in Supplementary Material~\cite{Suppl}. Here we only give two examples illustrating this criterion: for instance, for $\To=2.15$~K and $a=0.1$~cm the temperature is practically uniform provided $\vo\ll95$~cm/s; for the typical hot-wire radius $a=1\,\mu{\rm m}$ in Grenoble experiment, for $\To=2.15$~K the bulk temperature can be regarded as uniform provided $\vo\ll950$~cm/s). We also note that the values of $\vo$ for which the temperature is uniform become much larger as both $\To$ and $a$ decrease.

We have solved numerically (subject to the obvious condition $T=\To$ on $x=1$) the resulting nonlinear, first order, ordinary differential equation~(\ref{eq:T(x)}), in which the densities $\rhon$ and $\rhos$ (and their temperature derivatives), thermodynamic properties, and mutual friction parameters are known functions of temperature. The numerical task presented a certain problem because $\rhon$, $\rhos$, $s$, $c_s$, and $\alpha$ are known from experimental data only on discrete sets of temperature [$\gamma(T)$ can also be evaluated from numerical simulations~\cite{Adachi}]. With the exception of $\gamma$, the recommended values for these properties are given in Ref.~\cite{Donnelly-Barenghi}; for example, the normal and superfluid densities are tabulated~\cite{Donnelly-Barenghi} for discrete values of $T$ with steps $\Delta T=0.05\,{\rm K}$ for $0\leq T\leq2.1\,{\rm K}$, $\Delta T=0.01\,{\rm K}$ for $2.1\leq T\leq2.17\,{\rm K}$, and $\Delta T=0.001\,{\rm K}$ for higher temperatures (for other properties the temperature steps may be different). However, a numerical procedure for solving Eq.~(\ref{eq:T(x)}) requires that the properties listed above can be determined for any desired temperature, that is, their values are needed as continuous functions of temperature. This is particularly important because large temperature gradients in the region adjacent to the 
cylinder's surface and the numerical evaluation of the superfluid density derivatives in Eq.~(\ref{eq:F-C}) require a rather fine discretization of the $x$-axis.

To determine the values of $\rhon$, $\rhos$, $s$, $c_s$, and $\alpha$ at any desired temperature, we have used cubic spline fits of the properties of liquid helium recommended in Ref.~\cite{Donnelly-Barenghi} for discrete sets of temperature. The evaluation routine for splines is described in detail in the Appendix of the cited paper\footnote{Evaluated by cubic splines, the properies of liquid helium for any desired temperature can be found using our interactive website www.mas.ncl.ac.uk/helium/.}. Unlike $\rhon$, $\rhos$, $s$, $c_s$, and $\alpha$, the parameter $\gamma$ was measured or calculated for just a few values of temperature; for example, the values of $\gamma$ found from numerical simulation~\cite{Adachi} are reported for $T=1.3$, 1.6, 1.9, and 2.1~K. Solving numerically Eq.~(\ref{eq:T(x)}), we have simply interpolated the data~\cite{Adachi} using the cubic polynomials\footnote{Note that it is not yet possible to estimate an accuracy of this interpolation which, however, is well within the spread of available experimental and numerical data summarized in Ref.~\cite{Kondaurova}.}
\begin{equation}
10^{-3}\gamma(T)=0.3076T^3-1.621T^2+2.942T-1.708\,.
\label{eq:gamma-cubic}
\end{equation}

Eq.~(\ref{eq:T(x)}) was solved by means of the Adams-Bashforth method using the constant step $\Delta x=0.001$. At each step, the densities, thermodynamic properties, and the mutual friction coefficient 
were evaluated by means of cubic spline routines. The temperature derivative of the superfluid density in Eq.~(\ref{eq:F-C}) was calculated at each $x$ from the cubic spline interpolation of $\rhos(T)$. A direct check showed that further decreasing $\Delta x$ does not affect the numerical accuracy of the solution for any surface temperature $\To<2.16\,{\rm K}$ (but much smaller $\Delta x$ would be 
required for higher $\To$).

\begin{figure}[h]
\begin{center}
    \includegraphics[width = 0.6\linewidth]{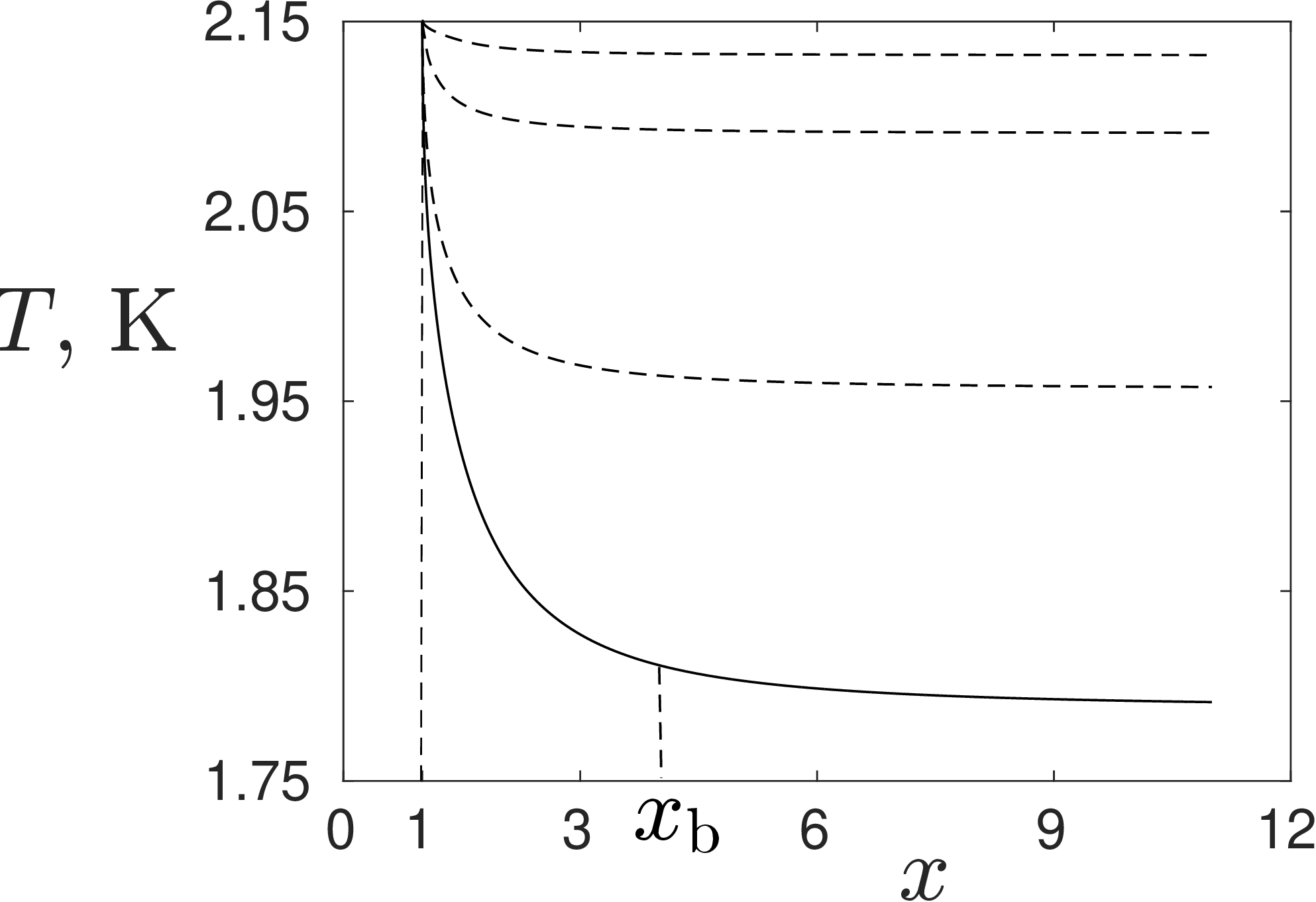}
    \caption{Radial distribution of temperature around the cylinder of 0.1\,cm radius for surface temperature $\To=2.15\,{\rm K}$. Solid line shows $T(x)$ for $\vo=60\,{\rm cm/s}$, with $x_{\rm b}$ indicating the position of thermal boundary layer. Dashed lines, from bottom to top, correspond to $\vo=45$, 20, and 10~cm/s.}
    \label{fig:fig1}
\end{center}
\end{figure}

\begin{figure}[htb]
\centering
  \begin{tabular}{@{}cc@{}}
    \includegraphics[width = 0.53\linewidth]{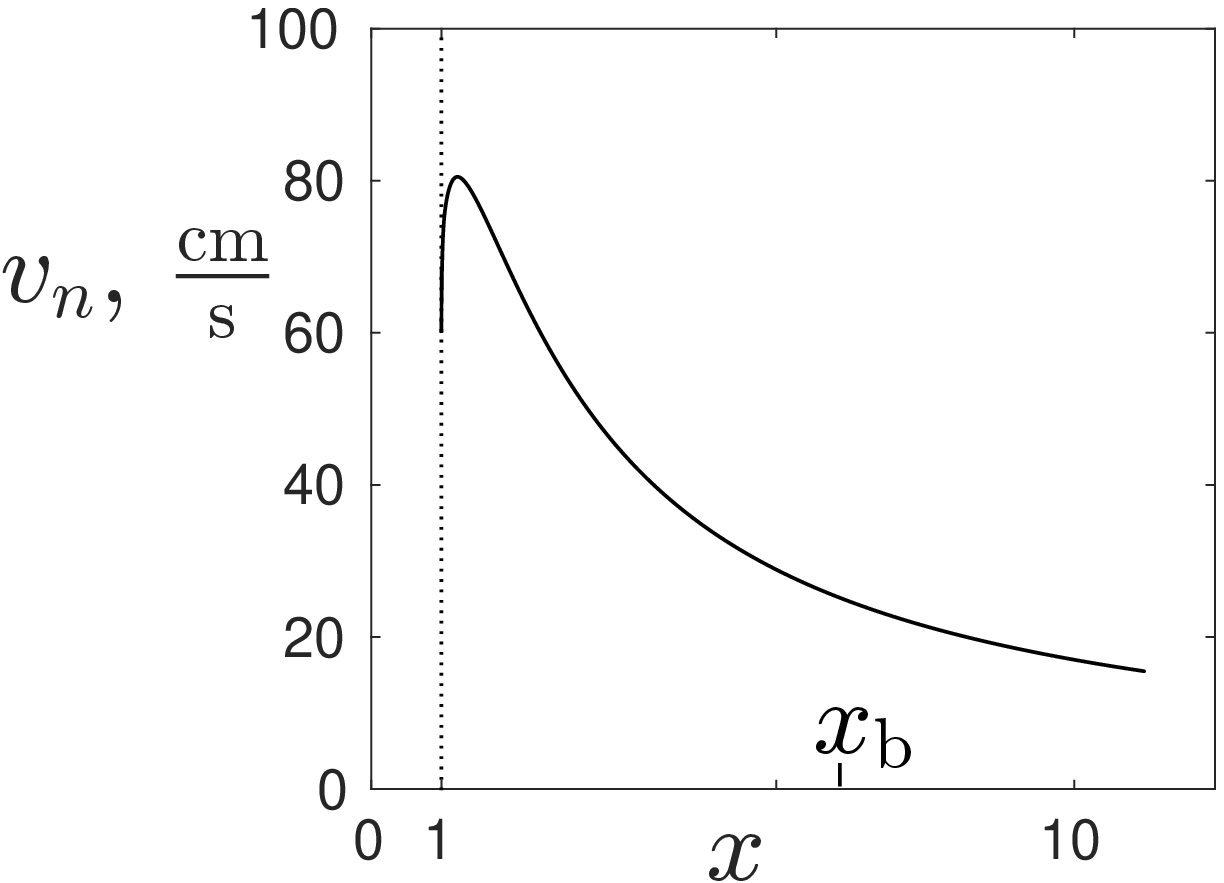} &
    \includegraphics[width = 0.40\linewidth]{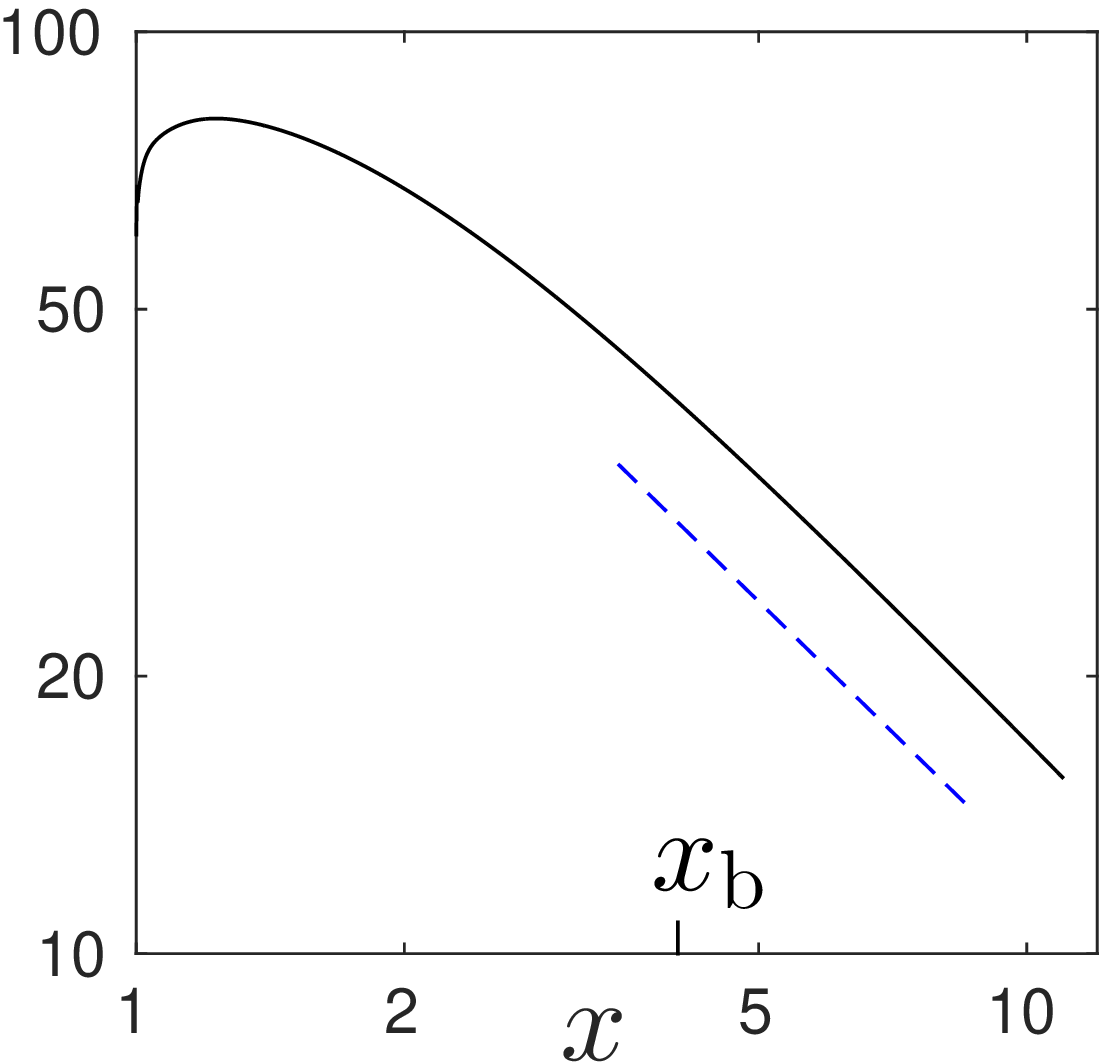} \\
\end{tabular}
    \caption{Radial distribution of the normal velocity, $\vn$ for $\To=2.15\,{\rm K}$ and $\vo=60\,{\rm cm/s}$ (heat flux $q=28\,{\rm W/cm^2}$). Right panel shows $\vn(x)$ vs $\log x$; $x^{-1}$ scaling is shown by the dashed line.}
    \label{fig:fig2}
\end{figure}

\begin{figure}[htb]
\centering
  \begin{tabular}{@{}cc@{}}
    \includegraphics[width = 0.51\linewidth]{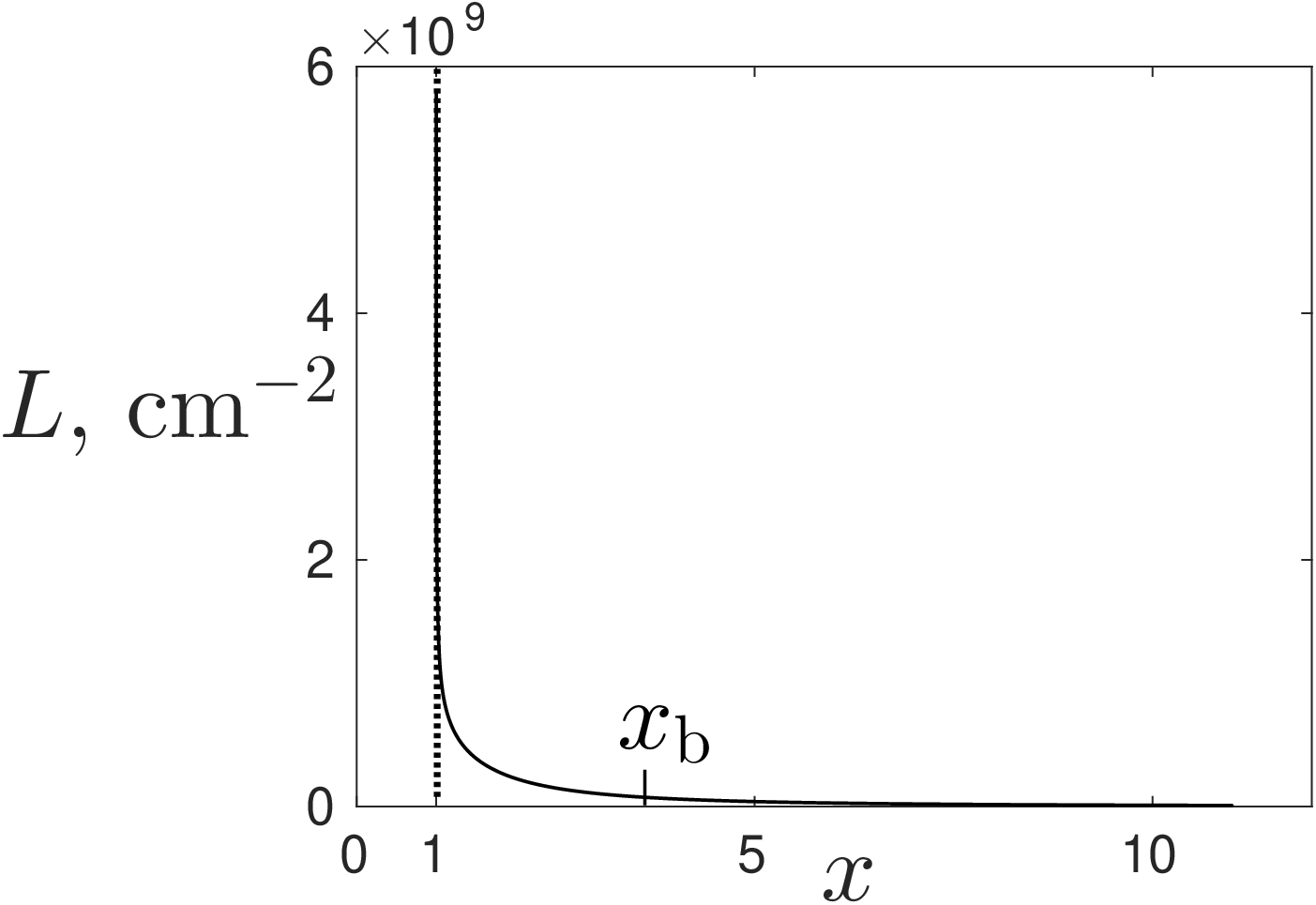} &
    \includegraphics[width = 0.40\linewidth]{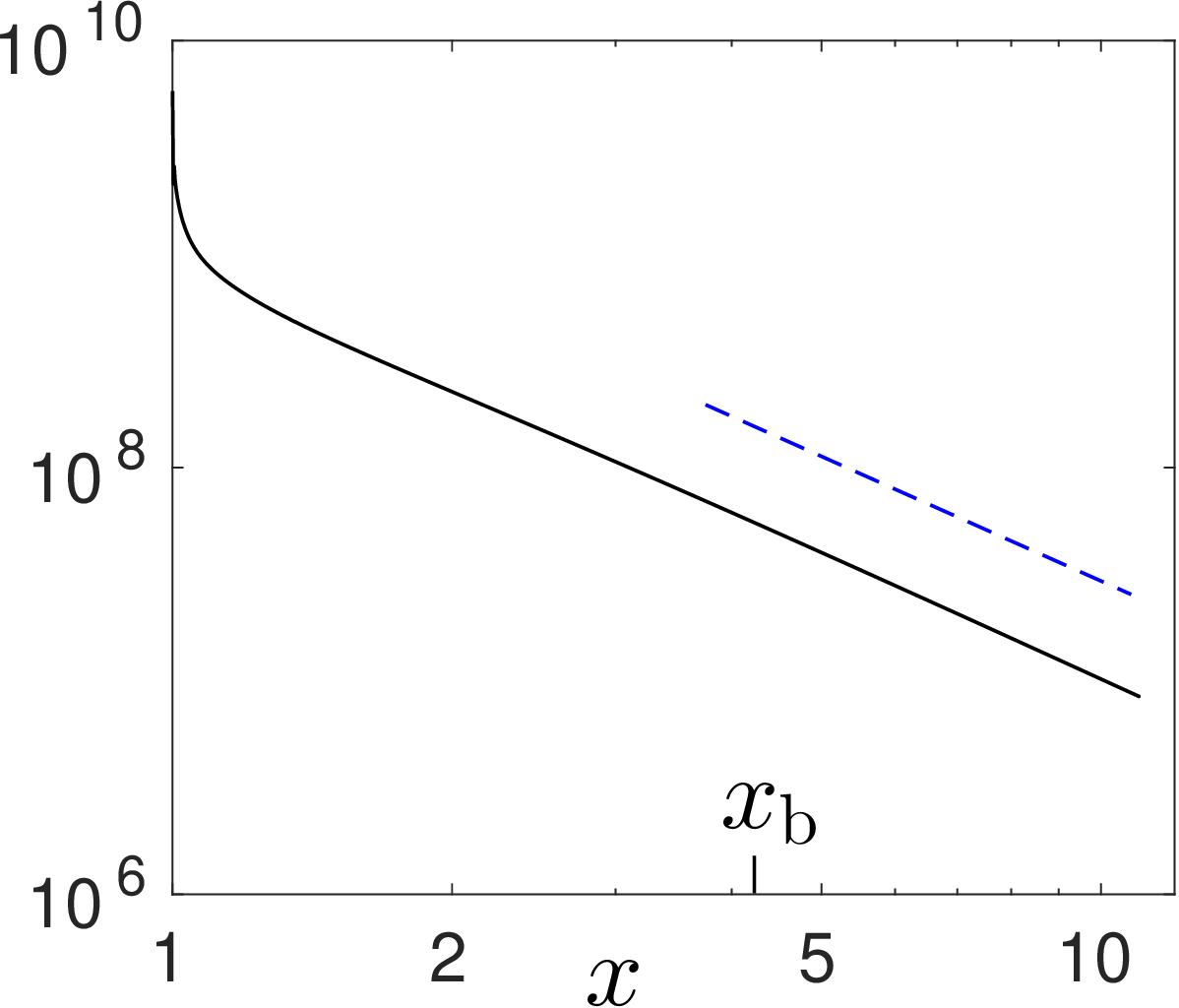} \\
\end{tabular}
    \caption{Radial profile of the vortex line density (the values of parameters are the same as in Fig.~\ref{fig:fig2}). Right panel shows $L(x)$ vs $\log x$, with the $x^{-2}$ scaling shown by the dashed line.}
    \label{fig:fig3}
\end{figure}

Unlike the uniform-temperature solution, a non-isothermal, steady-state solution exists for any surface temperature $\To<\Tl$. A typical behavior of temperature, normal and counterflow velocities, 
and local vortex line density with the distance from the center of cylinder is illustrated in Figs.~\ref{fig:fig1}-\ref{fig:fig3}. As the distance from the cylinder increases, $T(x)$ tends
to a constant temperature, $T_\infty$ fully determined by the surface temperature, $\To$ and the velocity $\vo$ (that is, by the heat flux $q$). In other words, the steady-state solution exists only for a single bath temperature, $T_{\rm bath}=T_\infty(\To,\,\vo)$. To analyze a radial counterflow in the general case where $\To$ and $\vo$ are imposed at the same time for a given $T_\infty$ would require finding a time-dependent solution of the HVBK equations which, perhaps, should include the second order transfer processes (viscous dissipation, etc.) neglected in the current approach. Such a problem presents formidable difficulties and is outside the scope of this work which aims at investigating the flow properties, temperature, and vortex line density distributions in the relatively close vicinity of the cylinder where a very strong counterflow dominates other mechanisms of heat transfer.

We now describe the results of our calculations in more detail. Figures~\ref{fig:fig1}-\ref{fig:fig3} show the radial distributions, $T(x)$, $\vn(x)$ and $L(x)$ for the cylinder of radius $a=0.1\,{\rm cm}$, the surface temperature $\To=2.15\,{\rm K}$, and the normal velocity at the surface $\vo=60\,{\rm cm/s}$, the latter corresponding to the heat flux $q=27\,{\rm W/cm^2}$ (note that the heat flux in the hot-wire experiments~\cite{Shiotsu,Ruzhu,Duri} could be from the order of several tens up to $250\,{\rm W/cm^2}$). The values of parameters~(\ref{eq:beta}) are $\beta=1.92\times10^{-4}\,{\rm K}$ and $\sigma=2.6\times10^{-2}\,{\rm K}$.

Illustrating the temperature distribution, Fig.~\ref{fig:fig1} points to the formation of cylindrical shell within which the temperature sharply drops from $\To=2.15\,{\rm K}$ to the temperature close to its asymptotic value $T_\infty=T_{\rm bath}\approx1.79\,{\rm K}$. For simplicity we call this shell a ``thermal boundary layer''\footnote{This, relatively thick layer is not a classical boundary layer whose thickness is determined by a competition between the convection of heat along the surface and the thermal conduction in the direction normal to the surface. As we did not account for the molecular conductivity, the mechanism of formation of the thermal boundary layer in our work is somewhat different from that in Ref.~\cite{Duri}.} and associate, rather arbitrarily, its outer boundary with the value of $x=x_{\rm b}$ such that $T(x_{\rm b})-T_\infty=0.01T_\infty$. In Fig.~\ref{fig:fig1}, the solid line corresponds to $\vo=60\,{\rm cm/s}$; in this case $x_{\rm b}\approx4.2$ so that the non-dimensional thickness  of the thermal boundary layer is $\delta=x_{\rm b}-1\approx3.2$ (in dimensional units $\delta_{\rm d}=a\delta\approx320\,\mu{\rm m}$). Temperature distributions for $T=2.15\,{\rm K}$ and smaller values of $\vo$ are shown in Fig.~\ref{fig:fig1} by dashed lines which indicate that with the decrease of the heat flux the temperature distribution becomes more shallow: with $\vo$ decreasing below 10~cm/s the tempearture reaches its asymptotic value $T_\infty=T_{\rm bath}$ at smaller and smaller distances from the surface so that the boundary layer eventually becomes negligible. This behavior remains qualitatively the same for other values of $\To$ and $a$, see below for further details.

Having found the radial profile of temperature (and, hence, the radial distributions of the normal and superfluid densities, thermodynamic properties, $\alpha$, and $\gamma$) we then infer the radial distributions of the velocities from relations~(\ref{eq:v-rad}) and of the vortex line density from Eq.~(\ref{eq:gamma}).

Figure~\ref{fig:fig2} shows the radial profile of the normal velocity for $\To=2.15\,{\rm K}$ and $\vo=60\,{\rm cm/s}$. An interesting feature is that the normal velocity increases with $x$ at small distances from the heated surface, reaching the peak of $\vn\approx80\,{\rm cm/s}$ at some $x$ within the thermal boundary layer. Outside the boundary layer (for $x>x_{\rm b }$) the normal velocity decreases with distance as $x^{-1}$, as should be expected for the isothermal radial counterflow. The behavior of the counterflow velocity, $\vns$ (not shown) is somewhat less interesting: it sharply decreases at very small distances from the surface and quickly acquires the $x^{-1}$ behavior outside the boundary layer.

Our calculation shows the formation of the very dense vortex tangle, with the value of local line density $L\approx5.8\times10^9\,{\rm cm^{-2}}$ in the immediate vicinity of the cylinder's surface, see Fig.~\ref{fig:fig3} (note that local vortex line densities of a similar order of magnitude were expected~\cite{Duri} in the close vicinity of the hot-wire). Within a distance about the cylinder's radius from the surface the vortex line density decreases by more than an order of magnitude to $L\approx2\times10^8\,{\rm cm^{-2}}$ and then follows, within the remaining part of the thermal boundary layer and beyond, the scaling $L\sim x^{-2}$ typical of the isothermal, radial, turbulent counterflow.

For all simulations reported above and below, we estimated, from our numerical solutions, the contribution of the viscous stress term, $\eta\nabla^2\vvn$ to the momentum transfer in the normal fluid. We found that, even in the immediate vicinity of the cylinder's surface ($x=1$) where the derivatives of temperature and other properties are very large, the magnitude of the viscous stress term remains much smaller (by at least an order of magnitude) than the magnitude of the mutual friction force $-\rhos\alpha\kappa L\vvns$. This justifies the omission of the viscous term from the HVBK equations in our version of the model\footnote{Note that the inclusion of the viscous stress term into the momentum flux density tensor~(\ref{eq:tensor}), although leading to the second-order equation with a small parameter in front of $d^2T/dx^2$ [instead of the first-order ordinary differential equation~(\ref{eq:T(x)})] would not yield a classical boundary layer-type solution for $T(x)$ because in this case the second boundary condition, $T=T_{\rm bath}$ should be accommodated at large distances from the cylinder.}. Likewise, having estimated the individual terms of Eq.~(\ref{eq:Vinen}) we found that the convective term $\bnabla\cdot(L{\bf v}_L)=\bnabla\cdot(b\vvns)$ in the left-hand-side of this equation remains smaller by at least an order of magnitude than both the production and the destruction terms in the right-hand-side. This justifies the reduction of Eq.~(\ref{eq:Vinen}) to the Gorter-Mellink form~(\ref{eq:gamma}).

\begin{figure*}[htb]
\centering
  \begin{tabular}{@{}ccc@{}}
    \includegraphics[width = 0.33\linewidth]{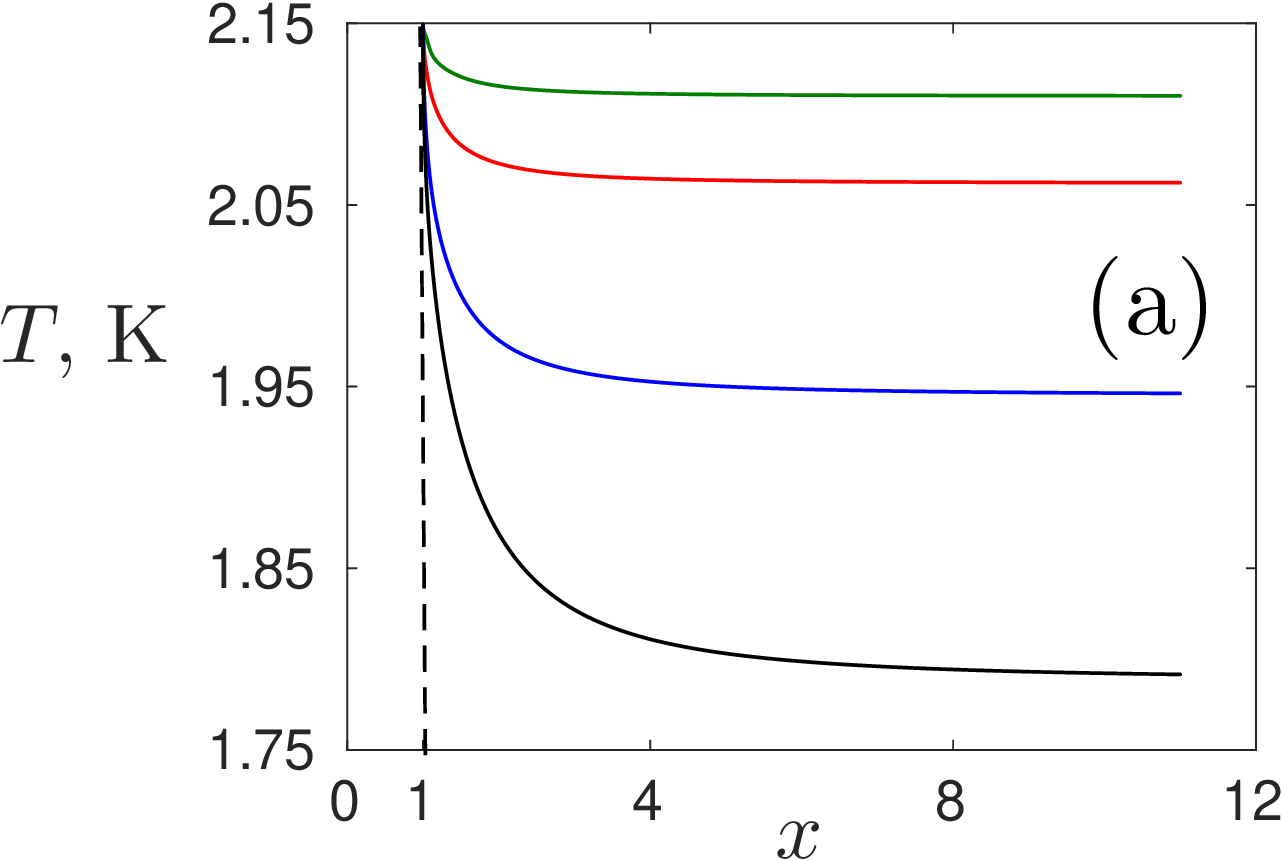} &
    \includegraphics[width = 0.28\linewidth]{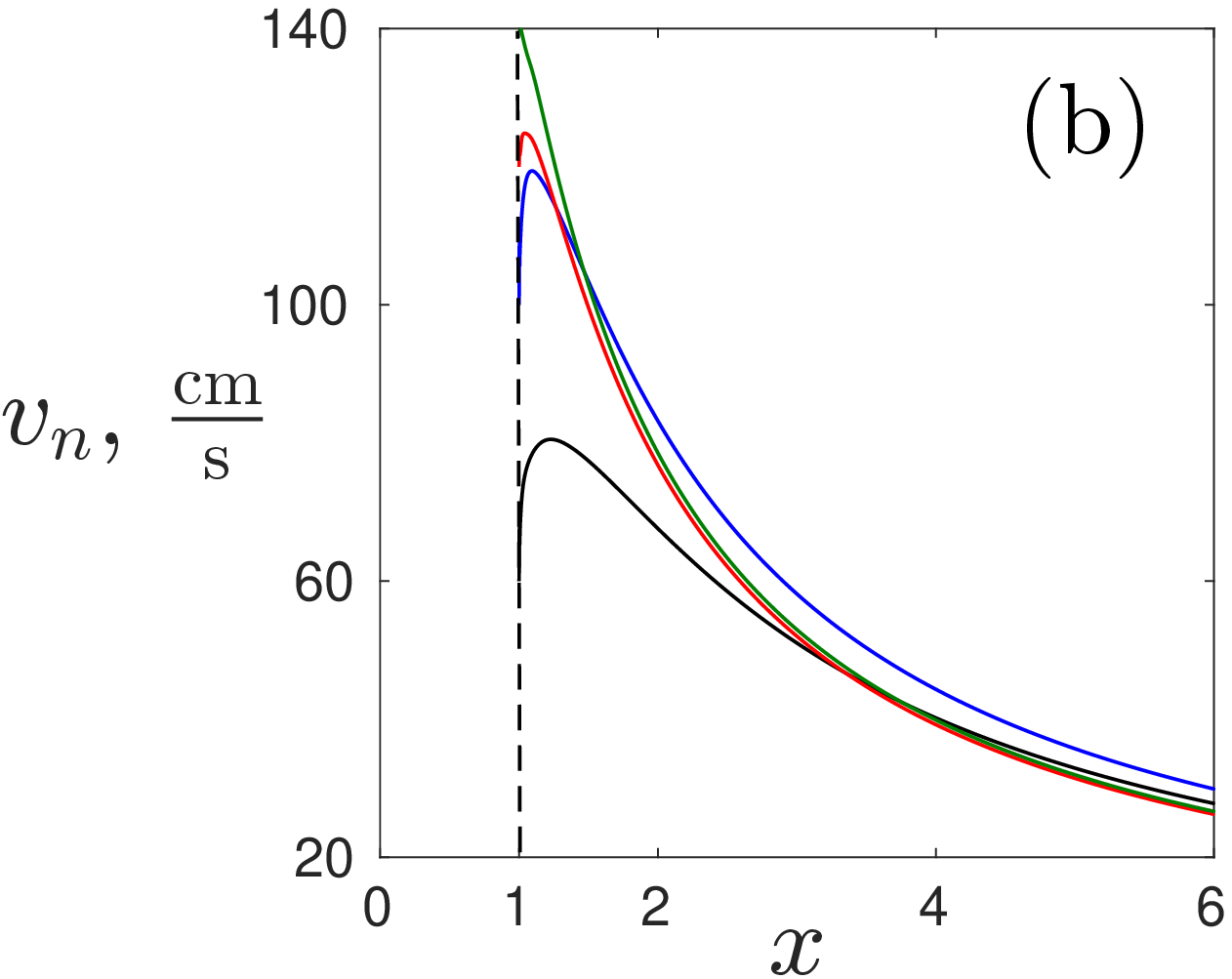} &
    \includegraphics[width = 0.28\linewidth]{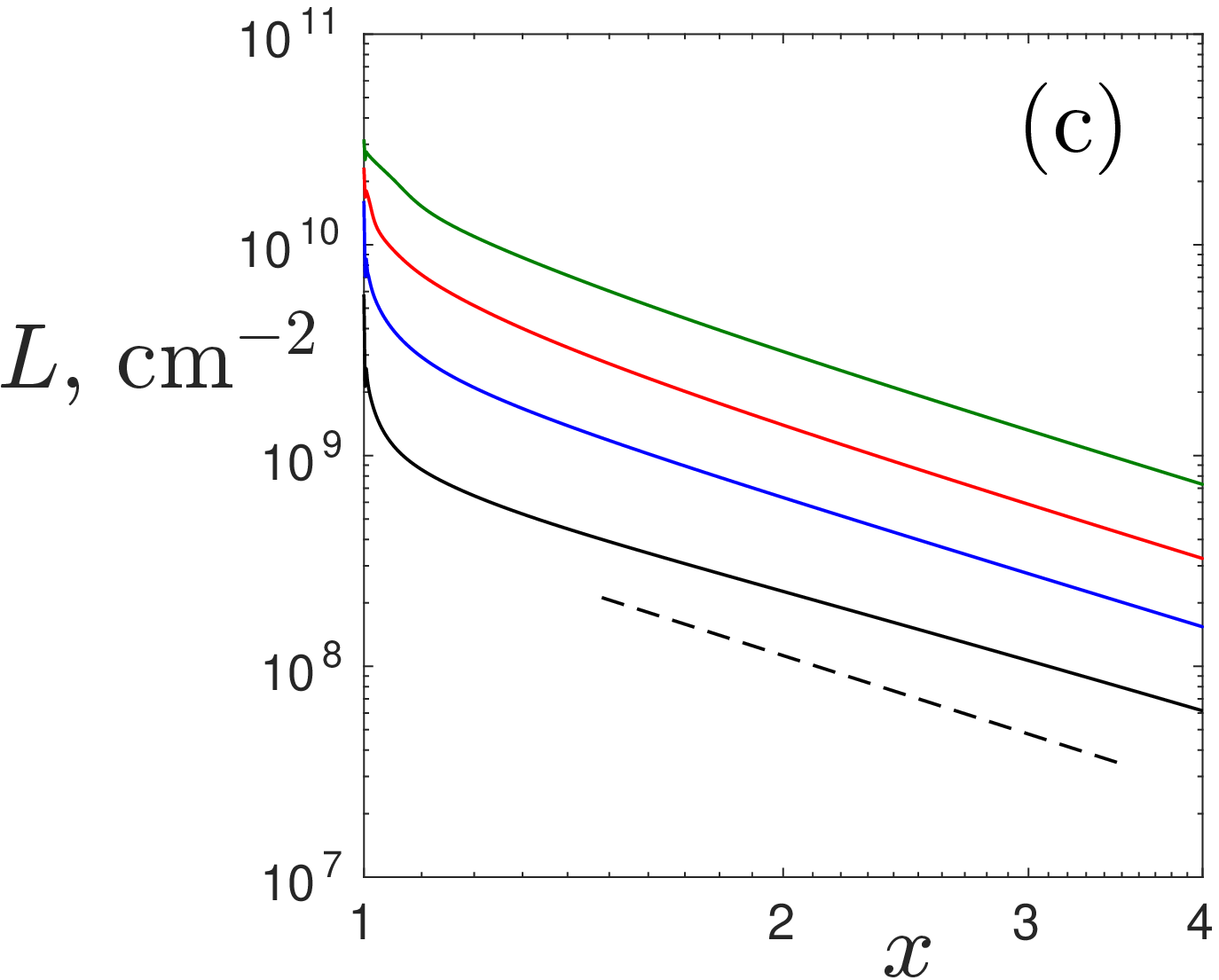} \\
\end{tabular}
    \caption{Radial profiles of (a) temperature, (b) normal velocity, and (c) vortex line density for $\To=2.15\,{\rm K}$ and various velocities $\vo$ and cylinder radii $a$. Black: $a=0.1\,{\rm cm}$, $\vo=60\,{\rm cm/s}$; blue: $a=100\,\mu{\rm m}$, $\vo=100\,{\rm cm/s}$; red: $a=10\,\mu{\rm m}$, $\vo=120\,{\rm cm/s}$; green: $a=1\,\mu{\rm m}$, $\vo=140\,{\rm cm/s}$. The dashed line in panel~(c) shows the $x^{-2}$ scaling.}
    \label{fig:fig4}
\end{figure*}

For surface temperature $\To=2.15\,{\rm K}$, we illustrate below how the radial profiles of temperature, normal velocity, and vortex line density are affected by the cylinder's radius and the velocity $\vo$ (that is, by the heat flux $q$). Panels (a), (b), and (c) of Fig.~\ref{fig:fig4} show
\begin{figure*}[htb]
\centering
  \begin{tabular}{@{}ccc@{}}
    \includegraphics[width = 0.28\linewidth]{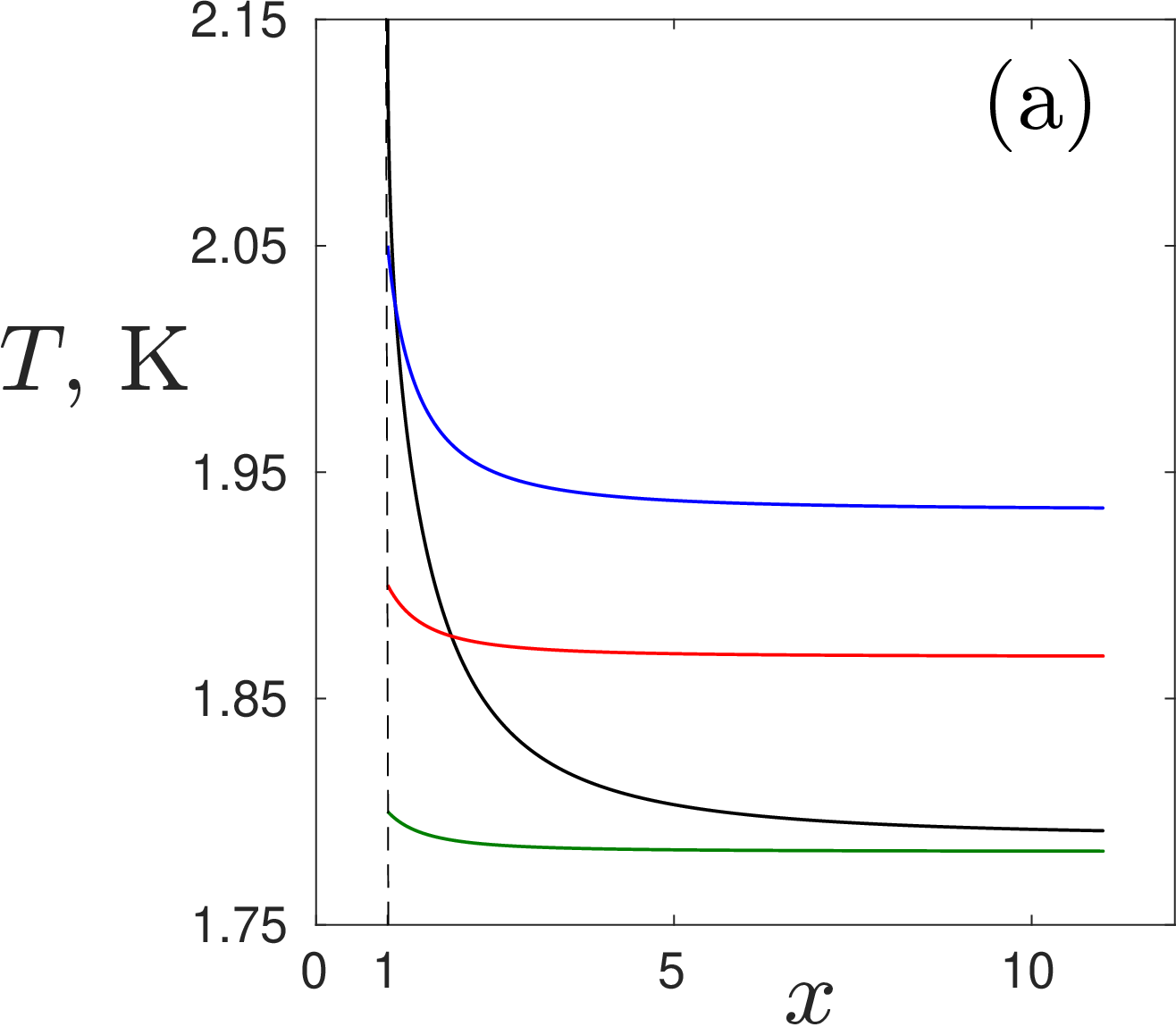} &
    \includegraphics[width = 0.31\linewidth]{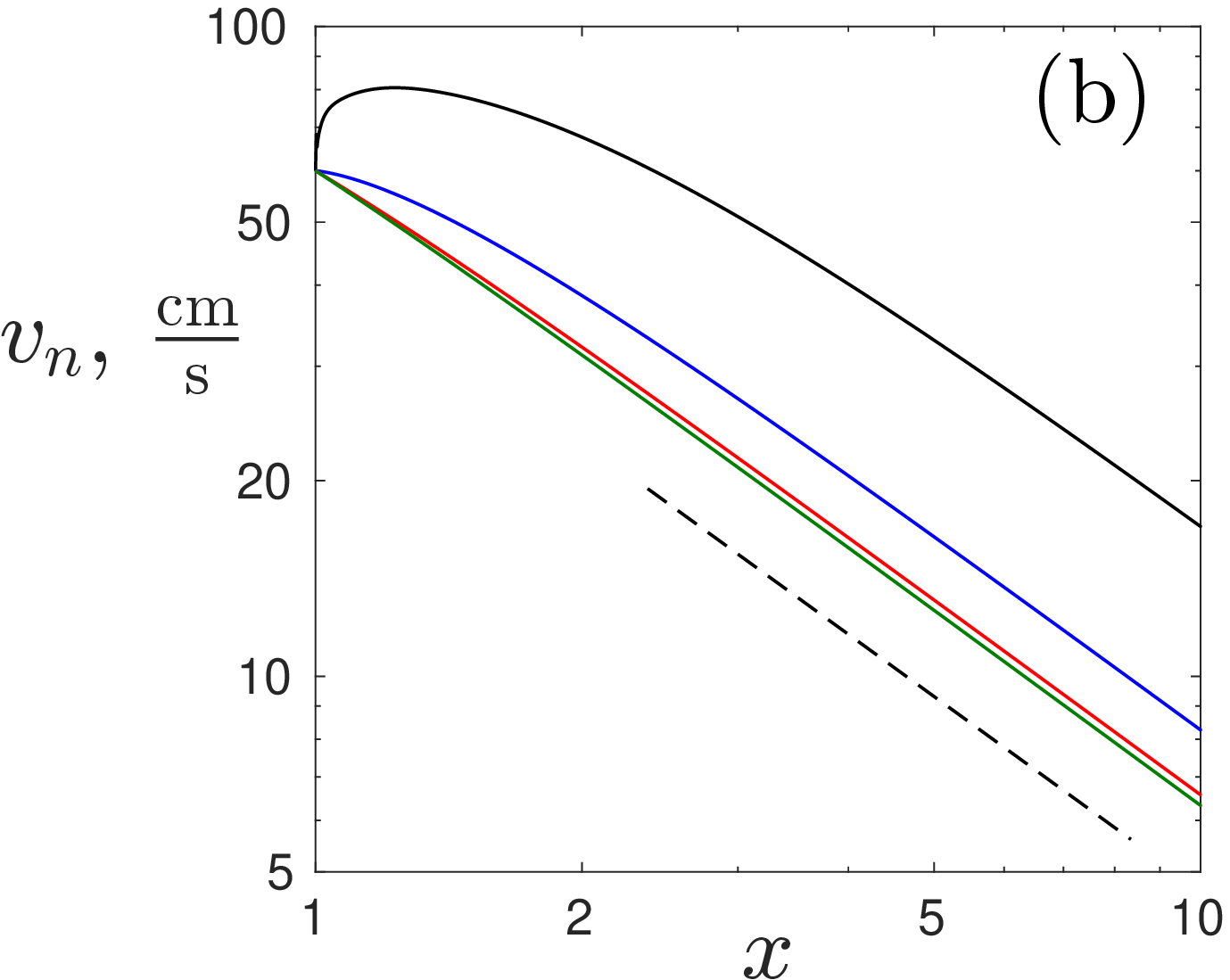} &
    \includegraphics[width = 0.30\linewidth]{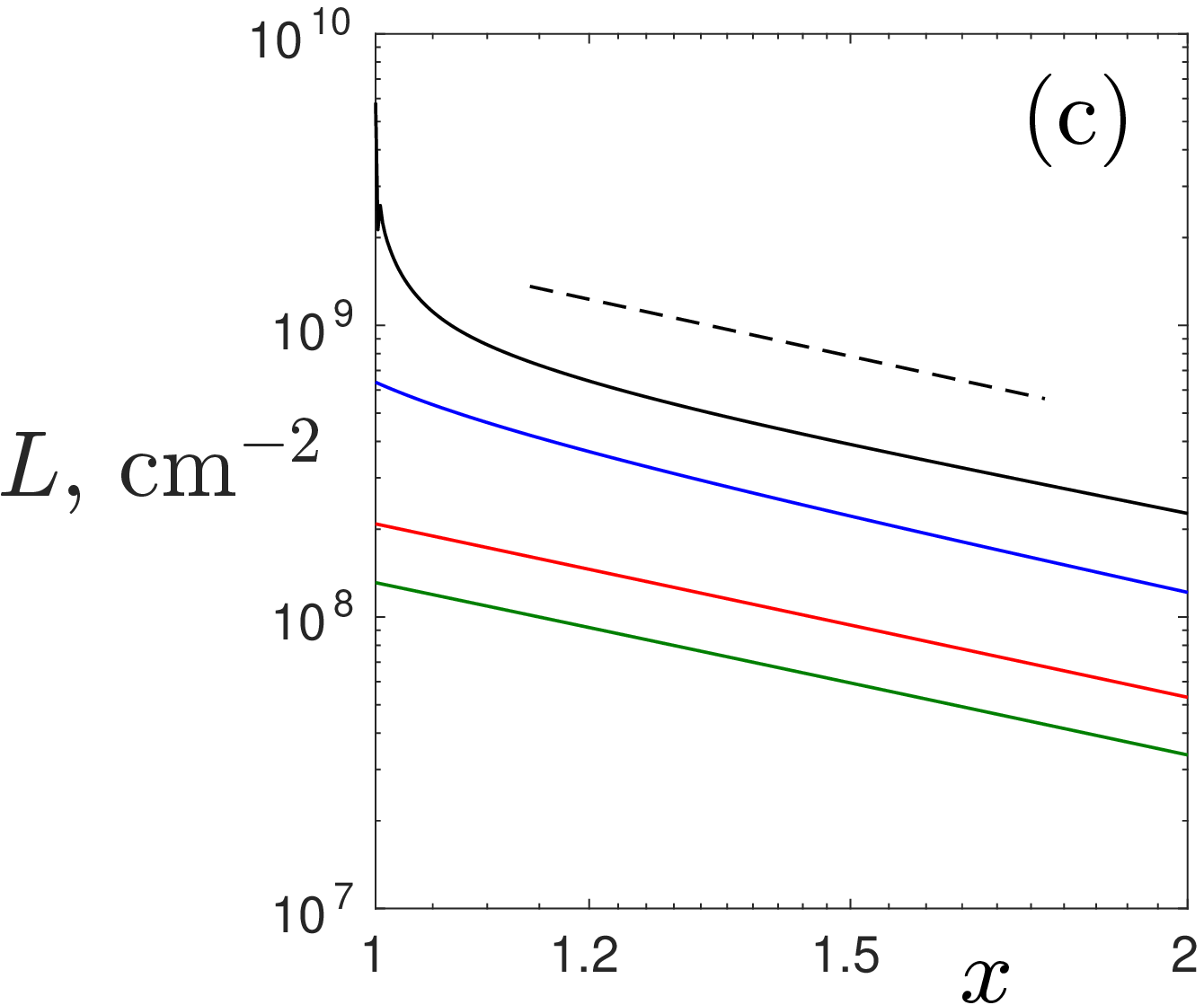} \\
\end{tabular}
    \caption{Radial distributions, for $a=0.1\,{\rm cm}$ and $\vo=60\,{\rm cm/s}$, of (a) temperature, (b) normal velocity, and (c) vortex line density for surface temperatures $\To=2.15\,{\rm K}$ (black), 2.05~K (blue), 1.9~K (red), and 1.8~K (green). In panels (b) and (c) these surface temperatures correspond to curves from top to bottom. The dashed lines in panels~(b) and (c) show the $x^{-1}$ and $x^{-2}$ scalings, respectively.}
    \label{fig:fig5}
\end{figure*}
$T(x)$, $\vn(x)$, and $L(x)$, respectively. All distributions are qualitatively similar to those described in more detail earlier for $a=0.1\,{\rm cm}$ and $\vo=60\,{\rm cm/s}$ (although the peak of the normal velocity near the cylinder's surface becomes indiscernible for $a=1\,\mu{\rm m}$ and $\vo=140\,{\rm cm/s}$). Clearly, the non-dimensional thickness of the thermal boundary layer depends on $\To$, $a$, and $\vo$; this will be discussed later.

Figure~\ref{fig:fig5} illustrates, for $a=0.1\,{\rm cm}$ and 
$\vo=60\,{\rm cm/s}$ the influence of surface temperature, $\To$ on the radial profiles of temperature, normal velocity, and vortex line density. As expected, for fixed values of $a$ and $\vo$ the thickness of the boundary layer decreases with a decrease of $\To$. For $\To\leq1.8\,{\rm K}$ the thermal boundary layer is no longer discernible so that the steady-state solution is practically the same as the isothermal distributions of $\vn(x)$ and $L(x)$ with $T=\To=T_{\rm bath}$, and, in dimensional variables, the normal, superfluid, and counterflow velocities are well approximated by formulae~(\ref{eq:v-rad}) in which $\rhon=\rhon(\To)$, $\rhos=\rhos(\To)$, $s=s(\To)=\so$, and the vortex line density profile is $L=A/r^2$, where $A=a^2\rho^2\vo^2\gamma^2(\To)/\rhos^2(\To)$.

\begin{figure}[htb]
\centering
  \begin{tabular}{@{}ccc@{}}
    \includegraphics[width = 0.29\linewidth]{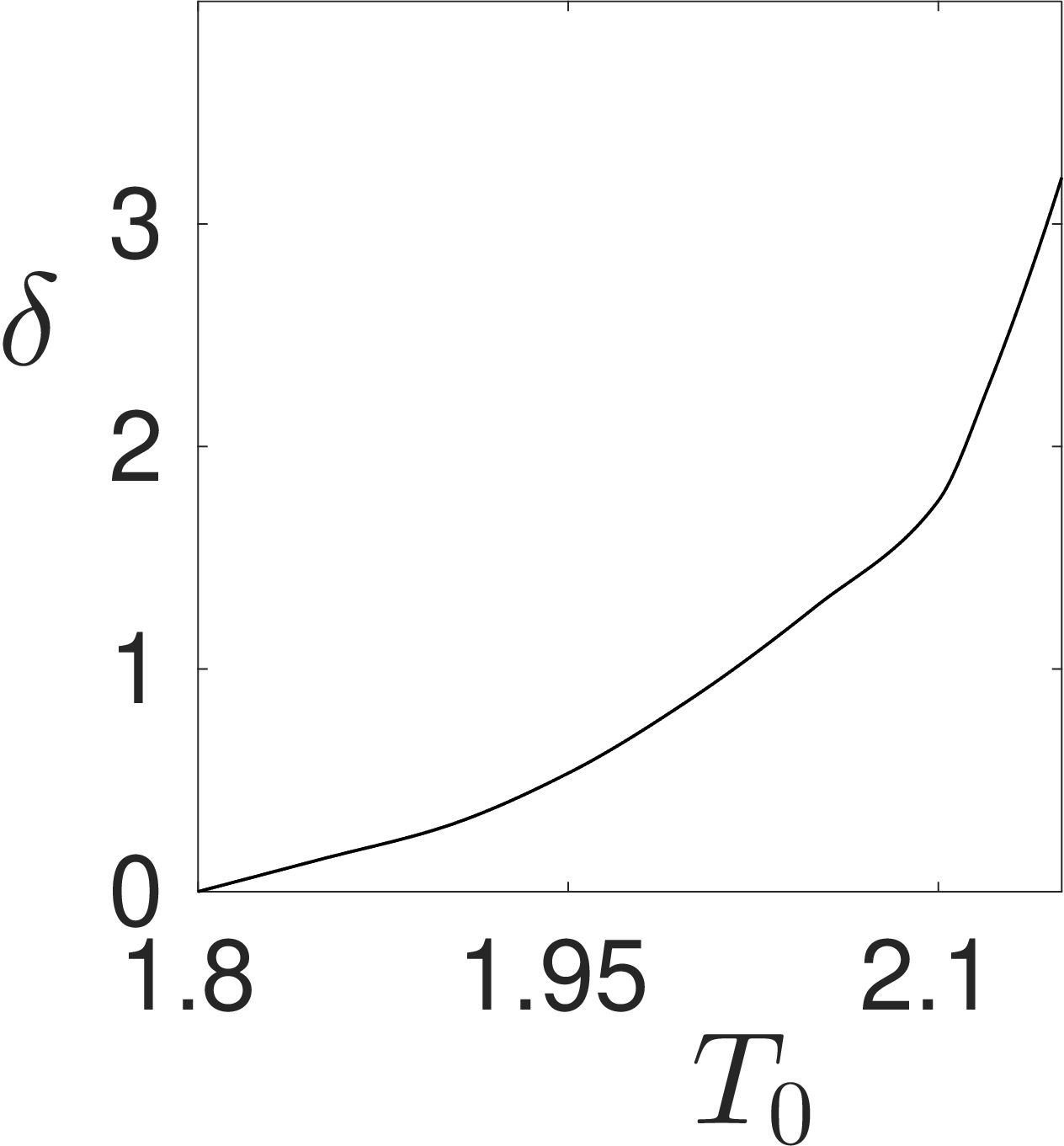} &
    \includegraphics[width = 0.282\linewidth]{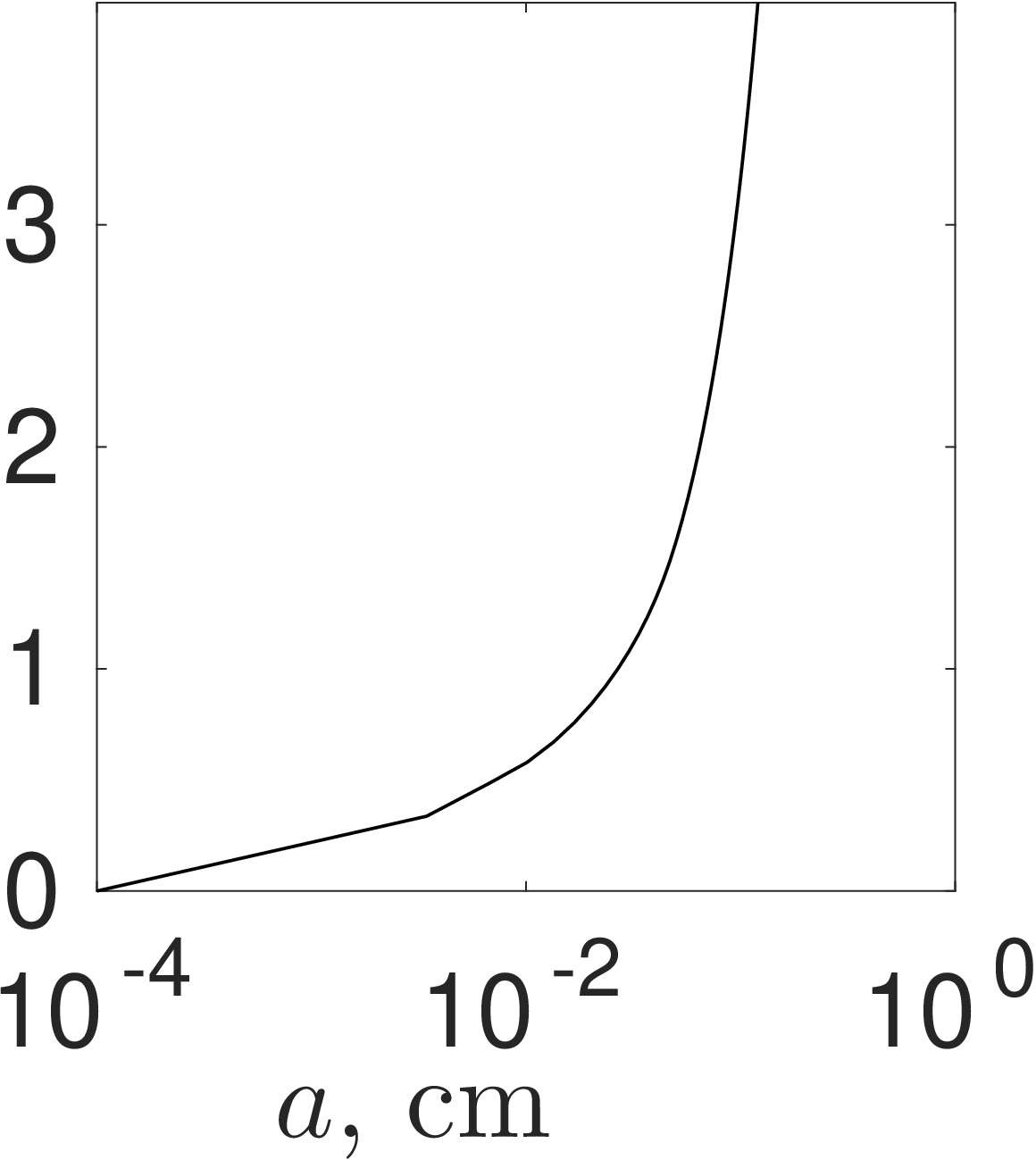} &
    \includegraphics[width = 0.282\linewidth]{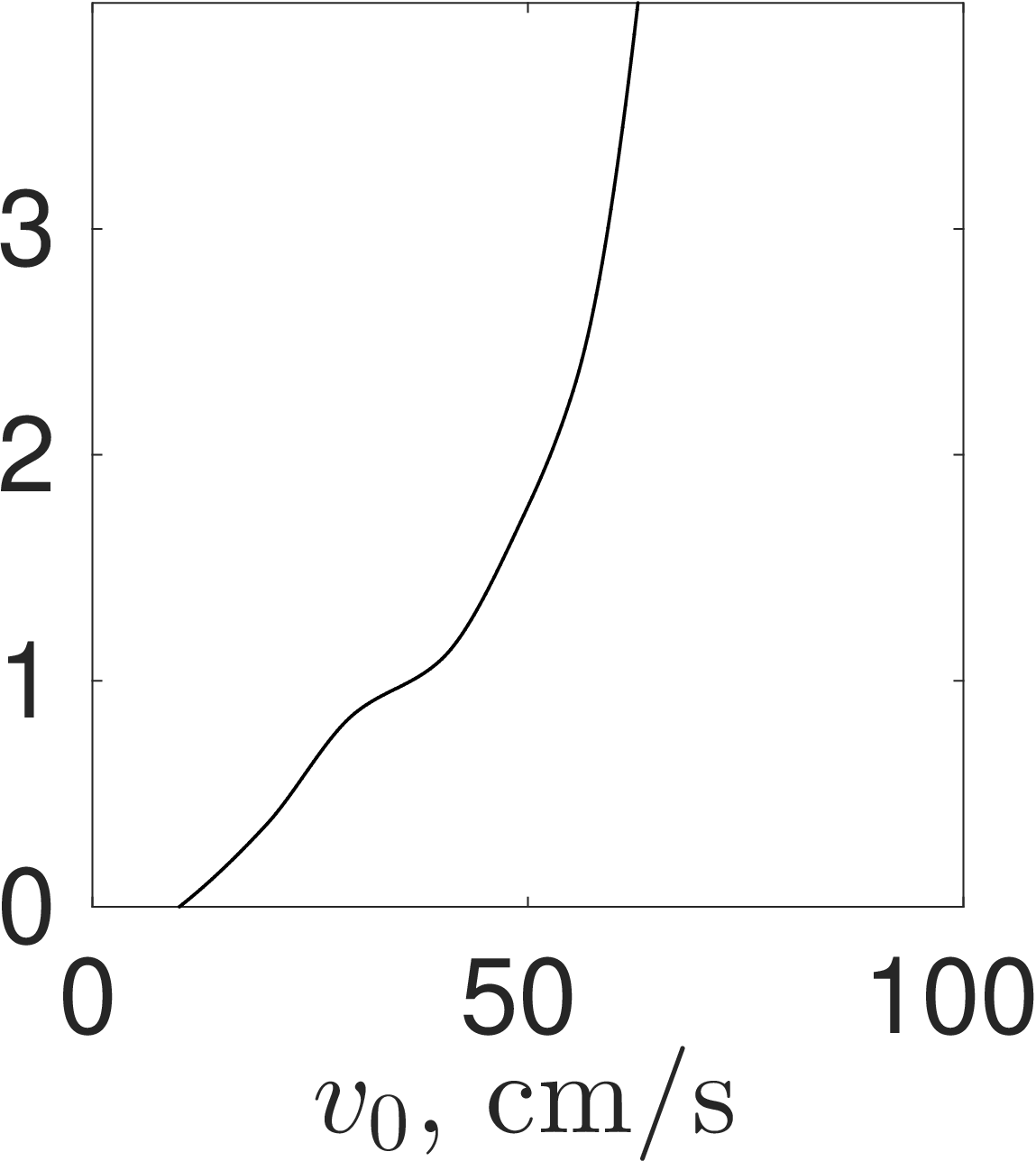} \\
\end{tabular}
    \caption{Non-dimensional thickness of the thermal boundary layer vs surface temperature $\To$ for $a=0.1\,{\rm cm}$, $\vo=60\,{\rm cm/s}$ (left); cylinder radius $a$ for $\To=2.15\,{\rm K}$, $\vo=60\,{\rm cm/s}$ (center); and velocity $\vo$ for $\To=2.15\,{\rm K}$, $a=0.1\,{\rm cm}$ (right).}
    \label{fig:fig6}
\end{figure}

Finally, in Fig.~\ref{fig:fig6} we show how the non-dimensional thickness,
$\delta$ of thermal boundary layer is affected by the surface temperature $\To$, cylinder radius $a$, and the velocity $\vo$ (that is, by the heat flux $q$). As we already noted earlier, the thermal boundary layer becomes negligible in the case where either the heat flux is small or the surface temperature is lower than 1.8~K.

\section{Conclusions}
\label{sec:conclusions}

We applied the HVBK model to the analysis of two-dimensional, steady state, radial turbulent counterflow generated by the heated cylinder immersed in $^4$He. We found that a time-independent solution of the HVBK equations can be found only if a spatial non-uniformity of helium temperature and the dependence on temperature of the normal and superfluid densities, thermodynamic properties, and mutual friction parameters are accounted for. Our numerical solutions showed the formation of the thermal boundary layer, whose thickness grows from practically zero at $\To=1.8\,{\rm K}$ to several cylinder radii at $\To=2.15\,{\rm K}$, within which the temperature rapidly decreases with the distance from the cylinder's surface. This implies a rapid change with a radial distance, $r$ of the superfluid density and thermodynamic properties (entropy and heat capacity) which are very sensitive to temperature in the vicinity of the $\lambda$-point. A rapid change with $r$ of the temperature within the thermal boundary layer affects the radial distributions of the normal and superfluid velocities, and of the vortex line density. Outside the boundary layer the temperature remains practically constant so that the velocities' and the local vortex line density profiles scale with the radial distance as $r^{-1}$ and $r^{-2}$, respectively, as should be expected for the constant-temperature radial turbulent counterflow. At surface temperatures below 1.8~K, or/and low heat fluxes from the cylinder's surface, the thermal boundary layer becomes negligible and the temperature remains practically constant throughout the entire flow domain.

Data supporting this publication is openly available under an Open Data Commons Open Database License~\cite{data}.

\begin{acknowledgments}
This work was partially supported by EPSRC grant EP/R005192/1.
\end{acknowledgments}


\begin{thebibliography}{99}
\bibitem{data}
DOI to be provided once accepted.
\end{thebibliography}
\end{document}


\title{Supplementary Material\\Turbulent radial thermal counterflow in the framework of the HVBK model}

\author{Y.\,A.~Sergeev}
\affiliation{Joint Quantum Centre Durham-Newcastle, and School of Mathematics, Statistics and Physics, Newcastle University, Newcastle upon Tyne, NE1 7RU, UK}

\author{C.\,F.~Barenghi}
\affiliation{Joint Quantum Centre Durham-Newcastle, and School of Mathematics, Statistics and Physics, Newcastle University, Newcastle upon Tyne, NE1 7RU, UK}

\maketitle

\subsection{Equation for the radial distribution of temperature}
A distribution of temperature in the radial, cylindrically symmetric, turbulent counterflow generated by the infinitely long heated cylinder of radius $a$, maintained at temperature $\To$, is governed by the equation
\begin{equation}
\left[1-\frac{\beta}{x^2}F(T)\right]\frac{{\rm d}T}{{\rm d}x}=\frac{1}{x^3}[\beta G(T)-\sigma H(T)]\,.
\label{eq:T(x)}
\end{equation}
In Eq.~(\ref{eq:T(x)}), the functions $F(T)$, $G(T)$, and $H(T)$ are
\begin{align}
&F(T)=\frac{1}{2}\frac{\rn}{\rs S^3}\left(\frac{1}{\rs}\frac{{\rm d}\rs}{{\rm d}T}+\frac{2C_s}{ST}\right)\,,
\label{eq:F-C} \\
&G(T)=\frac{\rn}{\rs S^3}\,, \quad H(T)=\frac{\alpha\Gamma^2}{S^4\rs^3}\,, \label{eq:GH}
\end{align}
and the parameters $\beta$ and $\sigma$, which have the dimension of temperature, are
\begin{equation}
\beta=\so^2\vo^2/\sll^3\,, \quad \sigma=\kappa\so^3\vo^3\gamma_{21}^2a/\sll^4\,,
\label{eq:beta}
\end{equation}
where $s$ is the specific entropy, $\so=s(\To)$, $\sll=1.583\times10^7\,{\rm erg/(g\,K)}$ is the entropy at the $\lambda$-point, $\vo$ is the normal velocity, at the cylinder's surface, determined by the heat flux $q$ generated by the heated cylinder through the well known relation
\begin{equation}
\vo=\frac{q}{\rho s_0\To}\,,
\label{eq:q}
\end{equation}
$\rho$ is the density of liquid helium, $\kappa=0.997\times10^{-3}\,{\rm cm^2/s}$ is the quantum of circulation, and $\gamma_{21}$ is the Gorter-Mellink parameter $\gamma(T)$ at temperature 2.1~K. In Eq.~(\ref{eq:T(x)}) and relations (\ref{eq:F-C})-(\ref{eq:GH}) the temperature $T$ is dimensional, but the radial coordinate, $r$, normal and superfluid densities ($\rhon$ and $\rhos$, respectively) and velocities ($\vn$ and $\vs$, respectively), the counterflow velocity $\vns$, specific entropy $s$, heat capacity $c_s$, and the coefficient $\gamma$ are scaled as follows:
\begin{align}
&x=\frac{r}{a}\,, \quad r_{n,s}=\frac{\rho_{n,s}}{\rho}\,, \quad V_{n,s,ns}=\frac{v_{n,s,ns}}{\vo}\,, \nonumber \\ 
&S=\frac{s}{\sll}\,, \quad C_s=\frac{c_s}{\sll}\,, \quad \Gamma=\frac{\gamma}{\gamma_{21}}\,;
\label{eq:non-dim}
\end{align}
$\alpha$ is the temperature-dependent, non-dimensional mutual friction coefficient.

The values of temperature-dependent properties $\rhon$, $\rhos$, $s$, $c_s$, and $\alpha$ have been tabulated in Ref.~\cite{Donnelly-Barenghi}; for the temperature dependence of $\gamma$ see e.g. Refs.~\cite{Adachi,Kondaurova}.

\subsection{Functions $F(T)$, $G(T)$, and $H(T)$}
\begin{figure*}
\centering
  \begin{tabular}{@{}ccc@{}}
    \includegraphics[width = 0.3\linewidth]{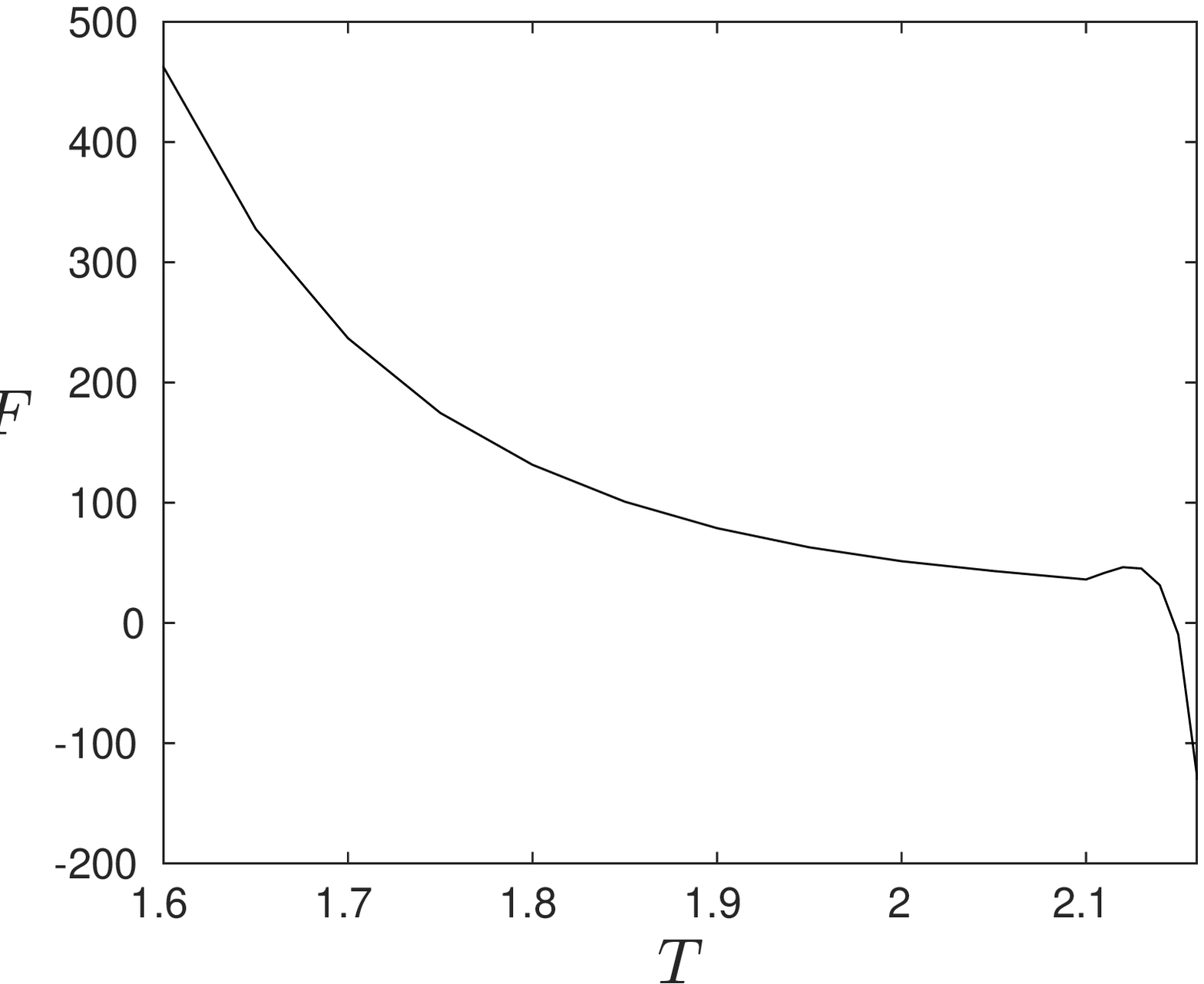} &
    \includegraphics[width = 0.28\linewidth]{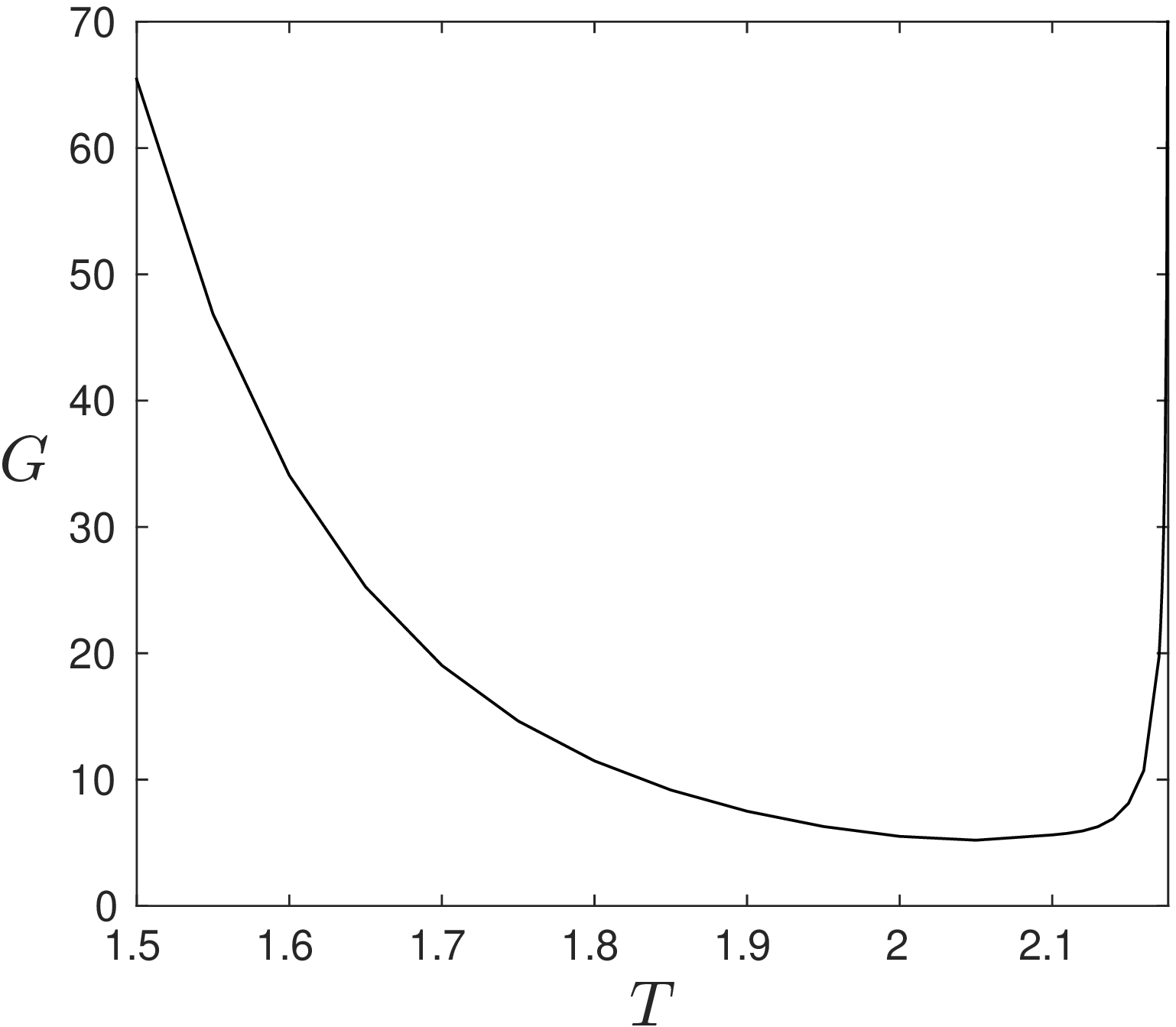} &
    \includegraphics[width = 0.26\linewidth]{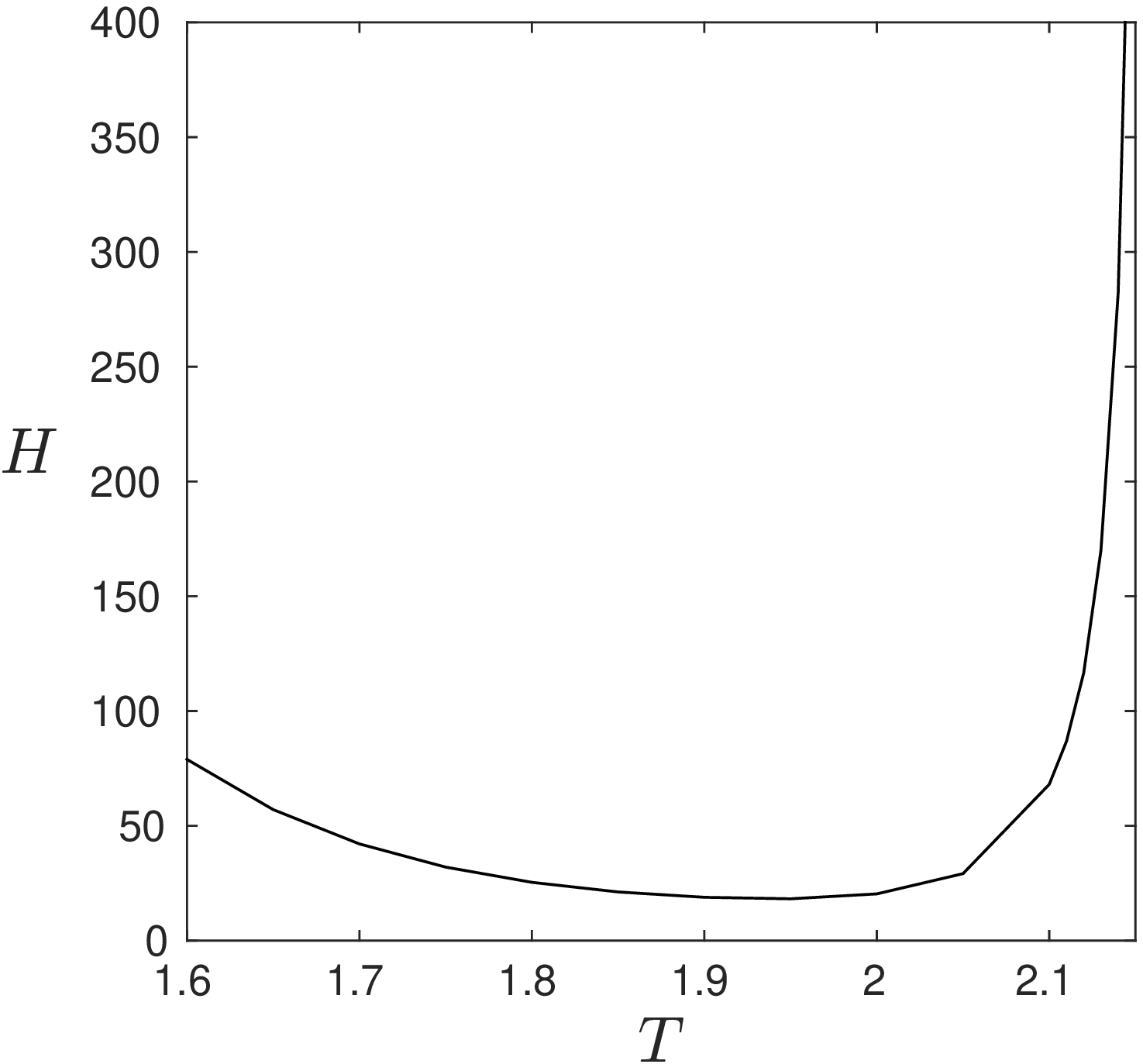} \\
\end{tabular}
\caption{Functions $F(T)$, $G(T)$, and $H(T)$.}
\label{fig:FGH}   
\end{figure*}

Functions $F(T)$, $G(T)$, and $H(T)$, defined by relations~(\ref{eq:F-C})-(\ref{eq:GH}), have been obtained using cubic spline fits of the properties of liquid helium recommended in Ref.~\cite{Donnelly-Barenghi} and interpolation of data~\cite{Adachi} for the parameter $\gamma$. These functions are plotted in Fig.~\ref{fig:FGH}.

We note here that for all physically meaningful values of parameters
\begin{equation}
1-\beta F(T)=O(1) ) \quad {\rm and} \quad \beta G(T)<\sigma H(T)
\label{eq:estimFGH}
\end{equation}
[in fact, in most cases $\beta G(T)\ll\sigma H(T)$].

\subsection{Hypothesis of uniform temperature: sufficient condition}
While we are mainly concerned with the distribution of temperature and flow properties in the case where the heat flux $q$ (and hence $\vo$) is relatively large and the surface temperature $\To$ is rather close to $\Tl$, intuitively it seems obvious that in the case where $\vo$ and $q$ are low, and/or the surface temperature $\To$ is not in the close vicinity of $\Tl$, the temperature of helium can be regarded as practically uniform throughout the whole flow domain.

\begin{figure}[h]
\begin{center}
    \includegraphics[width = 0.8\linewidth]{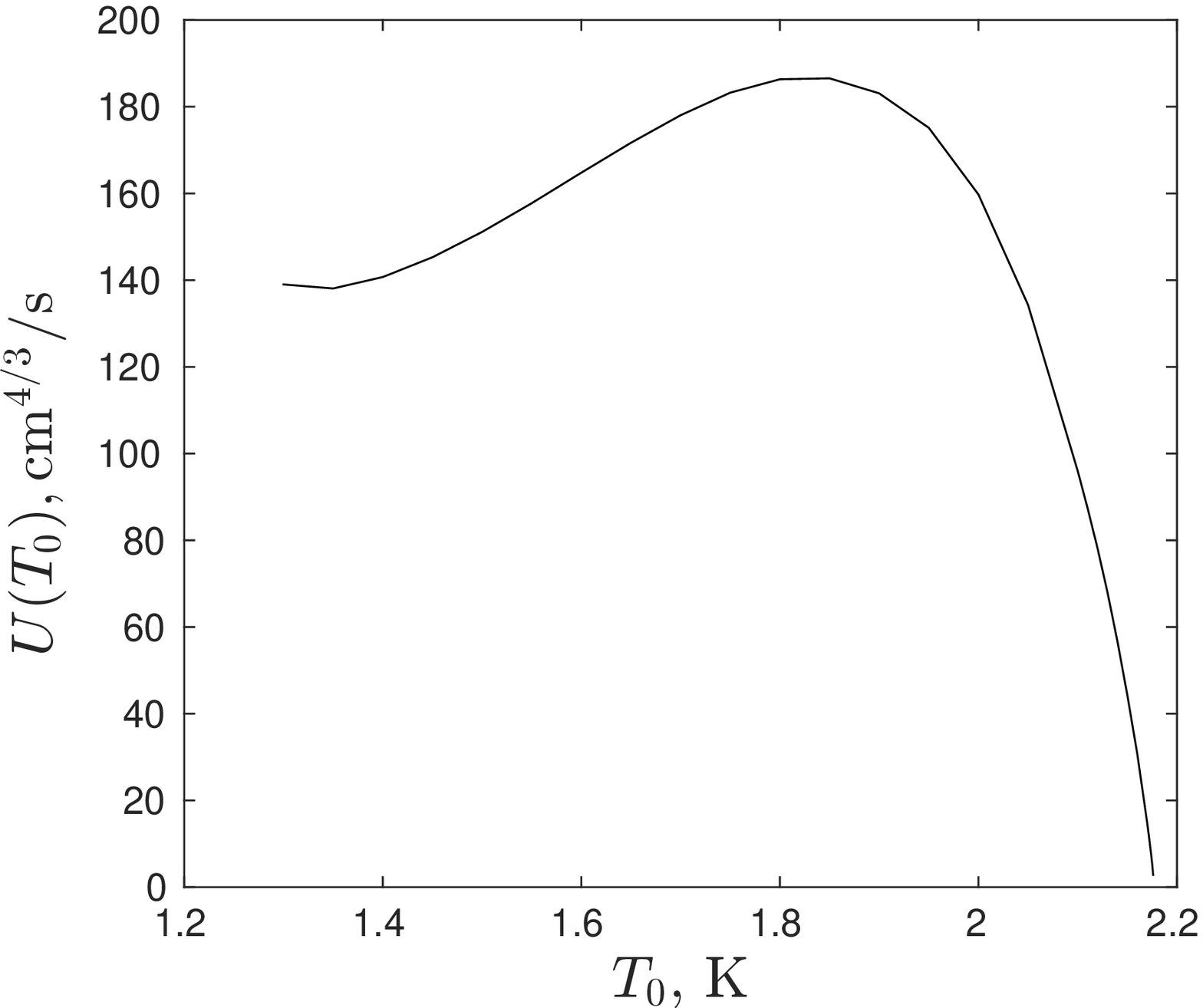}    
\end{center}
\caption{Function $U(\To)$ in the condition for the uniformity of temperature $a^{1/3}\vo\ll U(\To)$.}
\label{fig:U}
\end{figure}

To obtain the sufficient condition that would justify the hypothesis of uniform temperature, we consider Eq.~(\ref{eq:T(x)}). For all physically meaningful values of parameters the temperature decreases monotonically with $x$ (non-dimensional radial coordinate) so that the temperature can be considered practically uniform provided
\begin{equation} 
\frac{1}{\To}\left.\frac{{\rm d}T}{{\rm d}x}\right\vert_{x=1}\ll1\,,
\label{eq:dTdx}
\end{equation}
where $x=1$ corresponds to the surface of cylinder $r=a$. Then, making use of estimates~(\ref{eq:estimFGH}), we find that a sufficient condition for the uniformity of temperature can be formulated as $\sigma\To H(\To)\ll1$. Invoking Eqs.~(\ref{eq:beta}), this yields the condition in the form of the following inequality for normal velocity at the cylinder's surface:
\begin{equation}
a^{1/3}\vo\ll U(\To)\,,
\label{eq:condition-v}
\end{equation}
where
\begin{equation}
U(\To)=\frac{\rhos(\To)}{\rho}\left[\frac{\To s(\To)}{\kappa\alpha(\To)\gamma^2(\To)}\right]^{1/3}\,,
\label{eq:U}
\end{equation}
which, making use of relation~(\ref{eq:q}), can be reformulated as the condition for the heat flux
\begin{equation}
a^{1/3}q\ll\To^{4/3}s^{4/3}(\To)\frac{1}{[\kappa\alpha(\To)\gamma^2(\To)]^{1/3}}\,.
\label{eq:condition-q}
\end{equation}
Function $U(\To)$ is shown in Fig.~\ref{fig:U}.

A simple analysis of criteria~(\ref{eq:condition-v})-(\ref{eq:condition-q}) and Fig.~\ref{fig:U} shows that the values of $\vo$ and $q$, for which the temperature of helium can be regarded as uniform, become larger as $\To$ and $a$ decrease.

\subsection{Temperature distribution in the low heat flux limit}

In the limit of low heat flux (low $\vo$) the temperature gradients become small throughout the whole flow domain. In such a case, similarly to the customary approach to the analysis of thermal counterflow in channels, spatial variations of the normal and superfluid densities and thermodynamic and mutual friction parameters can be ignored, while the $\rhos s\bnabla T$ term should be kept in the HVBK (Hall-Vinen-Bekarevich-Khalatnikov) equations due to a large dimensional value of specific entropy.

Assuming for $\rhon$, $\rhos$, $s$, $\alpha$ and $\gamma$ the constant values corresponding to the temperature $T_\infty$ in the bulk of helium (that is, far from the heated cylinder), the HVBK equations yield the time-independent temperature distribution in the form
\begin{equation}
T(r)=T_\infty + \frac{\rho }{\rho_s} \frac{a^2}{r^2} \frac{v_0^2}{s} \left[ \frac{\rho ^3}{\rho_n \rho_s^2} \frac{\alpha \kappa \gamma^2 }{2} a v_0 -1 \right]\,.
\label{eq:T-smallgrad}
\end{equation}
Note that this immediately yields the condition for \textit{strict} uniformity of temperature:
\begin{equation}
a\vo=\frac{2}{\kappa\alpha\gamma^2}\frac{\rhon\rhos^2}{\rho^3}\,.
\label{eq:vo}
\end{equation}